\documentclass{article}
\usepackage{ijcai23}

\usepackage{times}
\usepackage{soul}
\usepackage{url}
\usepackage[hidelinks]{hyperref}
\usepackage[utf8]{inputenc}
\usepackage[small]{caption}
\usepackage{graphicx}
\usepackage{amsmath}
\usepackage{amsthm}
\usepackage{booktabs}
\usepackage{algorithm}
\usepackage{algorithmic}
\usepackage[switch]{lineno}

\newtheorem{theorem}{Theorem}

\usepackage{comment}
\usepackage{xspace}
\usepackage{amssymb} 

\newcommand{\Math}[1]{\ensuremath{#1}}

\newcommand{\modecal}[1]{{\Math{\mathcal{#1}}}}

\newcommand{\textmath}[1]{\mbox{\textit{#1}}}

\newcommand{\propername}[1]{\mbox{\small \textsf{#1}}\xspace}

\newcommand{\Golog}{\propername{Golog}}

\newcommand{\ConGolog}{\propername{ConGolog}}

\newcommand{\A}{\modecal{A}} 
\newcommand{\C}{\modecal{C}} \newcommand{\D}{\modecal{D}}
 \newcommand{\F}{\modecal{F}}

 \newcommand{\N}{\modecal{N}}

\newcommand{\ndet}{\mbox{$|$}}
\newcommand{\conc}{\mbox{$\parallel$}}

\newcommand{\Trans}{\textmath{Trans}}

\newcommand{\Poss}{\textmath{Poss}}

\def\planeaux!#1:#2<-#3!{\Math{#1 \mbox{\rm:} #2\; \leftarrow #3}}

\def\planeaux!#1<-#2!{\Math{#1 \leftarrow #2}}

\long\def\eatpar#1{%
\ifx#1\par                      
\let\nextmove=\eatpar           
\else
\let\nextmove=#1
\fi
\nextmove
}

\def\qed{\hfill{\qedboxempty}      
  \ifdim\lastskip<\medskipamount \removelastskip\penalty55\medskip\fi}

\def\qedboxempty{\vbox{\hrule\hbox{\vrule\kern3pt
                 \vbox{\kern3pt\kern3pt}\kern3pt\vrule}\hrule}}

\def\qedfull{\hfill{\qedboxfull}   
  \ifdim\lastskip<\medskipamount \removelastskip\penalty55\medskip\fi}

\def\qedboxfull{\vrule height 4pt width 4pt depth 0pt}

\newcounter{bean}

\newenvironment{tightenumerate}{
                \begin{list}{
                  {\mbox {
                      \arabic{bean}.\/}}}{\usecounter{bean}
                      \setlength{\itemsep}{-1pt}\setlength{\topsep}{0pt}}}{
                \end{list}}

\newenvironment{tightitemize}{
                \begin{list}{$\bullet$}{
                    \setlength{\itemsep}{-1pt}}{\setlength{\topsep}{0pt}}}{
                \end{list}}

\newcommand{\under}[1]{\mbox{\underline{\it\smash{#1}\vphantom{\lower.05ex\hbox{
x}}}}}

\newcommand{\commentarea}[1]{}

\newtheorem{definition}[theorem]{Definition}  
\newtheorem{proposition}[theorem]{Proposition}
\newtheorem{lemma}[theorem]{Lemma}     

\newtheorem{assumption}[theorem]{Assumption}
\newtheorem{constraint}[theorem]{Constraint}

\newcommand{\SitDet}{\mathit{SituationDetermined}}
\newcommand{\limp}{\supset}

\newcommand{\CGaxioms}{\mathcal{C}}

\newcommand{\anyseqhl}{\textsc{anyseqhl}}

\newcommand{\anyonehlon}{\textsc{any1hl}}

\title{Abstraction of Nondeterministic Situation Calculus Action Theories - Extended Version}

\author{
Bita Banihashemi$^1$
\and
Giuseppe De Giacomo$^2$\And
Yves Lesp\'{e}rance$^3$
\affiliations
$^1$Ronin Institute\\
$^2$University of Oxford\\
$^3$York University
\emails
bita@ronininstitute.org,
giuseppe.degiacomo@cs.ox.ac.uk,
lesperan@eecs.yorku.ca
}

\begin{document}

\maketitle

\begin{abstract}
    We develop a general framework for abstracting the behavior of
an agent that operates in a nondeterministic domain, i.e., where the
agent does not control the outcome of the nondeterministic actions,
based on the nondeterministic situation calculus and the \ConGolog
programming language.
We assume that we have both an abstract and a concrete
nondeterministic basic action theory,
and a refinement mapping which specifies how abstract
actions, decomposed into agent actions and environment
reactions, are implemented by concrete \ConGolog programs.
This new setting supports strategic
reasoning and strategy synthesis, by allowing us to quantify
separately on agent actions and environment reactions. We show that if
the agent has a (strong FOND) plan/strategy to achieve a goal/complete
a task at the abstract level, and it can always execute the nondeterministic abstract actions
to completion at the concrete level, then there exists a refinement of
it that is a (strong FOND) plan/strategy to achieve the refinement of
the goal/task at the concrete level.
\end{abstract}

\section{Introduction}
When working in realistic dynamic domains, the use of \emph{abstraction} has proven essential in many areas of artificial intelligence, for example, in improving the efficiency of planning (e.g., \cite{DBLP:conf/aips/ChenB21}), explaining agent's behavior (e.g., \cite{DBLP:conf/aips/SeegebarthMSB12}), and in reinforcement learning (e.g., \cite{DBLP:journals/ai/SuttonPS99}).





Recently, \cite{DBLP:conf/aaai/BanihashemiGL17} (BDL17) proposed a formal account of \emph{agent
abstraction} based on the situation calculus \cite{McCarthy1969:AI,Reiter01-Book} and the \ConGolog agent programming language \cite{DBLP:journals/ai/GiacomoLL00}.
They assume that one has a high-level/abstract action theory,
a low-level/concrete action theory, both representing the
agent's behavior at different levels of detail, and a \emph{refinement
mapping} between the two. 
The refinement mapping
specifies how each high-level action is implemented by a low-level
\ConGolog program and how each high-level fluent can be translated into
a low-level state formula. 
This work defines notions of abstractions
between such action theories in terms of the existence of a suitable
bisimulation relation \cite{DBLP:conf/ijcai/Milner71}
between their respective models. 
Abstractions have many useful properties that ensure
that one can reason about the agent's actions (e.g., executability,
projection, and planning) at the abstract level, and refine and
concretely execute them at the low level. The framework can also be
used to generate high-level explanations of low-level behavior.  

This framework was formulated assuming a deterministic environment, as usual in the situation calculus.  Hence, the only nondeterminism was coming from \ConGolog programs and was \emph{angelic} in nature, i.e., under the control of the agent \cite{DBLP:journals/jlp/LevesqueRLLS97,DBLP:journals/ai/GiacomoLL00}.

However, many agents operate in \emph{nondeterministic environments}
where the agent does not fully control the outcome of its actions  (e.g., flipping a coin where the outcome may be heads or tails).
Recently, \cite{DBLP:conf/kr/GiacomoL21} (DL21) proposed a simple and elegant situation calculus account of nondeterministic environments where they clearly  distinguish between the nondeterminism associated with agent choices and that associated with environment
choices, the first being angelic for the agent, and the second being devilish, i.e., not under the agent's control.\footnote{Earlier accounts dealing with the topic in the situation calculus were more complex and did not clearly distinguish between these two forms of nondeterminism, e.g., \cite{DBLP:journals/ijufks/PintoSSM00} which deals with stochastic actions, and \cite{DBLP:journals/ai/BacchusHL99} which deals with uncertainty, noisy acting, and noisy sensing.}
%
The presence of environment nondeterminism deeply influences reasoning about action: if the agent wants to complete a task, it cannot simply do standard logical reasoning (i.e., satisfiability/logical implication), but needs to do \emph{realizability/synthesis}, i.e.,  devise a strategy that, in spite of the uncontrollable reactions of the environment, guarantees successful completion of the task \cite{Chu63,PnueliR89,Abadi:Lamport:Wolper:1989}.
We see this, for example in planning, where in the classical deterministic setting the problem can be solved by heuristic search, while in fully observable nondeterministic (FOND) domains forms of adversarial (AND/OR) search are required  \cite{DBLP:conf/aips/CimattiRT98,DBLP:series/synthesis/2019Haslum}.

%
%
%

In this paper, we develop an account of agent abstraction in presence of nondeterministic
environments based on (DL21).
The refinement mapping between high-level fluents and low-level state formulas remains as in (BDL17).  Instead we consider each high-level action, now nondeterministic, as being composed of an \emph{agent action} and an (implicit) \emph{environment reaction}.  As a consequence, we map the agent action (without the environment reaction) into a low-level \emph{agent program} that appropriately reflects the nondeterminism of the environment, and the complete high-level action, including both the agent action and the environment reaction, into a low-level \emph{system program} that relates the high-level environment reaction to the low-level ones.

We show that the (BDL17) notion of $m$-bisimulation extends naturally to this new setting. This allows for exploiting abstraction when reasoning about executions as in (BDL17), in spite of the nondeterministic environment. But, notably, this new setting now supports strategic reasoning and strategy synthesis, by allowing us to quantify separately on agent actions and environment reactions. As a result, we can exploit abstraction by finding abstract strategies at the high-level and then refining them into concrete low-level ones.
In particular, we show that if
the agent has a (strong FOND) plan/strategy to achieve a goal/complete a task at the high level (i.e., no matter what the environment does), and it can always execute the nondeterministic high-level actions to completion at the low level (even if not controlling their outcome), then there exists a refinement of it that is a (strong FOND) plan/strategy to achieve the refinement of the goal/task at the low level.

\section{Preliminaries}\label{sec:backgroundCom}

\paragraph{Situation Calculus.}
 The \textit{situation calculus} is a well known predicate logic language
for representing and reasoning about dynamically changing
worlds \cite{McCarthy1969:AI,Reiter01-Book}. 
All changes to the world are the result of \textit{actions},
which are terms in the language. 
%
A possible world history is represented by a term called a
\emph{situation}. The constant $S_0$ is used to denote the initial
situation.
Sequences of
actions are built using the function symbol $do$, such that $do(a,s)$
denotes the successor situation resulting from performing action $a$
in situation $s$.
Predicates and functions whose value varies from situation to situation are
called \textit{fluents}, and are denoted by symbols taking a
situation term as their last argument (e.g., $Open(Door1,s)$).
The abbreviation  $do([a_1, \ldots, a_n],s)$ stands for
$do(a_n,do(a_{n-1},\ldots , do(a_1,s) \ldots ))$; 
for an action sequence $\vec{a}$, we often write $do(\vec{a}, s)$ for $do([\vec{a}],s)$.
%
%
%
In this language, a dynamic domain can be represented by a \emph{basic
  action theory (BAT)}, where successor state axioms represent the
causal laws of the domain and and provide a solution to the frame
problem \cite{Reiter01-Book}.
%
%
A special predicate $\Poss(a,s)$ is used to state that action $a$ is executable
in situation $s$; the precondition axioms 
characterize this predicate.
Abbreviation $Executable(s)$ means that every action performed in
reaching situation $s$ was possible in the situation in which it occurred.

\paragraph{Nondeterministic Situation Calculus.}
A major limitation of the standard situation calculus and BATs is that atomic
actions are deterministic.
\cite{DBLP:conf/kr/GiacomoL21} (DL21) propose a simple extension of
the framework to handle nondeterministic actions while preserving the
solution to the frame problem.
For any primitive action by the agent in a nondeterministic domain, there can be a number of different outcomes. (DL21) takes
the outcome as being determined by the agent's action
and the environment's reaction to this action. This is modeled by having every action type/function $A(\vec{x}, e)$ take an
additional environment reaction parameter $e$, ranging over
a new sort \emph{Reaction} of environment reactions. The agent
cannot control the environment reaction, so it performs the
reaction-suppressed version of the action $A(\vec{x})$ and the environment then selects a reaction $e$ to produce the complete
action $A(\vec{x}, e)$. 
We call the reaction-suppressed version of the action $A(\vec{x})$ an \emph{agent action} and the full version of the
action $A(\vec{x}, e)$ a \emph{system action}. 

\paragraph{Nondeterministic Basic Action Theories (NDBATs).} These can be seen as a special kind of BAT, where every action function takes an environment reaction parameter, and moreover, for each agent action $A(\vec{x})$, we

\begin{itemize}

\item have its agent action precondition denoted by:
  \begin{small} $Poss_{ag}(A(\vec{x}), s) \doteq \phi^{agPoss}_{A}(\vec{x}, s);$ \end{small}%


\item have a reaction independence requirement, stating that the precondition for the agent action is independent of any environment reaction: 
\begin{small}  $\forall e. Poss(A(\vec{x}, e), s) \limp Poss_{ag}(A(\vec{x}), s);$  \end{small}%



\item have a \emph{reaction existence}
requirement, stating that if the precondition of the agent
action holds then there exists a reaction to it which makes
the complete system action executable: \begin{small} $Poss_{ag}(A(\vec{x}), s) \limp \exists e. Poss(A(\vec{x}, e), s). $ \end{small}%


\end{itemize}

\noindent
The above requirements \emph{must} be entailed by the action theory
for it to be an NDBAT.

A NDBAT
$\D$ is the union of the following disjoint sets: foundational, domain
independent, axioms of the situation calculus ($\Sigma$) as in
standard BATs, axioms describing the initial situation $D_{S_0}$ as in
standard BATs, unique name axioms for actions ($\D_{una}$) as in
standard BATs, successor state axioms (SSAs) describing how fluents change after \emph{system} actions are performed ($\D_{ssa}$), and \emph{system} action precondition axioms, one for each action type, stating when the complete system action can occur ($\D_{poss}$); these are of the form: \begin{small} $Poss(A(\vec{x},e), s) \equiv \phi^{Poss}_{A}(\vec{x},e, s).$ \end{small}%

	

\begin{figure}[htp]
	\centering
		\includegraphics[scale=0.65]{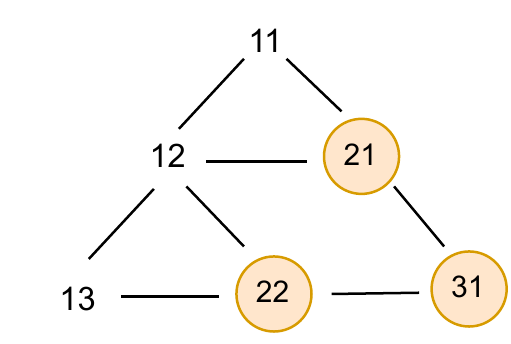}
	\caption{A 3-Level Triangle Tireworld}
	\label{fig:3LevTireWorld}
	\vspace*{-1.2em}
\end{figure}

\paragraph{Example.}
Our running example is based on a triangle tire-world domain (see Fig. \ref{fig:3LevTireWorld}). The agent's goal is to drive from location $11$ to location $13$. When driving from one location to the next, the possibility of a tire going flat exists. If there is a spare tire in the location where the car is at (indicated by circles in Fig. \ref{fig:3LevTireWorld}), the agent can use it to fix a flat. 


NDBAT $\D_l^{tt}$ models the domain at the low level.
The system action $drive(o,d,r)$ can be performed by the agent to drive from origin location $o$ to destination location $d$, 
and $r$ indicates environment's reaction. This action is executable
when the agent action $drive(o,d)$ is executable (i.e., when the agent
is at $o$, a road connects $o$ to $d$, and agent does not have a flat tire), and $r$ can take on two values: $\mathit{FlatTire}$ if the tire goes flat and $\mathit{NoFlatTire}$ otherwise.  
The fluent $At_{LL}(l,s)$ indicates agent's location, $Road_{LL}(o,d,s)$ specifies the road connections in Fig. \ref{fig:3LevTireWorld}, $\mathit{Flat_{LL}}(s)$ describes whether there is a flat tire, and $\mathit{Visited_{LL}}(l,s)$ indicates if the location $l$ has already been visited by the agent.


The system action $\mathit{fixFlatTire}(l,r)$ fixes a flat tire, and the
environment reaction $r$ is $Success_{LF}$ (assumed for
simplicity). It is executable if the fluent $Spare_{LL}(l,s)$ holds, i.e., there is a spare tire. The action $wait_{LL}$ can be performed by the agent to remain idle at a location.


$\D_l^{tt}$ includes the following  action precondition axioms (throughout the paper, we assume that free variables are universally quantified from the outside):

\noindent
\begin{small}
\[\begin{array}{l} 
Poss_{ag}(drive(o,d), s) \doteq  \\  
\hspace*{1.1em}  o \neq d \land At_{LL}(o,s) \land Road_{LL}(o,d,s) \land \neg \mathit{Flat_{LL}}(s) 
\\
Poss_{ag}(\mathit{fixFlatTire}(l), s) \doteq  \\
\hspace*{1.1em} At_{LL}(l,s) \land Spare_{LL}(l,s) \land \mathit{Flat_{LL}}(s)
\\
Poss_{ag}(wait_{LL}, s) \doteq TRUE
\\
Poss(drive(o,d,r), s) \equiv \\ 
\hspace*{1.1em}  Poss_{ag}(drive(o,d), s) \land (r=\mathit{FlatTire} \lor
    r=\mathit{NoFlatTire})
    \\

Poss(\mathit{fixFlatTire}(l,r), s) \equiv  \\
\hspace*{1.3em} Poss_{ag}(\mathit{fixFlatTire}(l), s) \land r=Success_{LF} 
\\
Poss(wait_{LL}(r), s) \equiv  \\
\hspace*{1.3em} Poss_{ag}(wait_{LL}, s) \land r=Success_{LW} 
\end{array} \]
\end{small}%
%


$\D_l^{tt}$ also includes the following SSAs:

\begin{small}
\[\begin{array}{l} 
At_{LL}(l, do(a,s)) \equiv 
\exists o,r. a = drive(o, l, r) \; \lor \\
\hspace*{1.3em} 
At_{LL}(l, s) \; \land \forall d,r. a \neq  drive(l, d, r)
\\
\mathit{Flat_{LL}}(do(a,s)) \equiv \\
\hspace*{1.5em} 
\exists o, d.  a=drive(o,d, \mathit{FlatTire}) \; \lor \\
\hspace*{1.5em}
\mathit{Flat_{LL}}(s) \land \forall r,l. a \neq \mathit{fixFlatTire}(l,r)
\\
\mathit{Visited_{LL}}(l,do(a,s)) \equiv \\
\hspace*{1.5em} 
\exists o, r.  a=drive(o,l, r) \; \lor \mathit{Visited_{LL}}(l,s)
\end{array} \]
\end{small}%

\noindent
For the other fluents, we have SSAs specifying that they are
unaffected by any action.


$\D_l^{tt}$  also contains the following initial state axioms:

\noindent
\begin{small}
\[\begin{array}{l} 
Road_{LL}(o,d,S_0) \equiv (o,d) \in \{ (11,12), (11,21), (12,21),\\
(12,22), (12,13),(13,22),(22,31),(21,31), (12,11), (21,11),\\
(21,12), (22,12), (13,12), (22,13),(31,22),(31,21) \}, \\
Spare_{LL}(l,S_0) \equiv l \in \{21,22,31\},  
\mathit{Visited_{LL}}(l,S_0) \equiv l=11, \\
At_{LL}(l,S_0) \equiv l=11, Dest_{LL}(l,S_0) \equiv l=13, \neg \mathit{Flat_{LL}}(S_0). 
\end{array} \]
\end{small}%

\paragraph{FOND Planning and Synthesis in NDBATs.}
We start with some definitions.
%
%
%
%
%
%
A \emph{weak plan} is one that achieves the goal
when the environment ``cooperates'' and selects environment reactions
to nondeterministic actions that make this happen.
Formally, we say that a sequence of agent actions $\vec{a}$ is a \emph{weak
plan} to achieve $Goal$ if $\exists s'. Do_{ag}(\vec{a},S_0,s') \land
Goal(s')$ holds,
i.e., there exists an execution of $\vec{a}$ that takes us from the
initial situation $S_0$ to situation $s'$ where  the goal holds.
$Do_{ag}(\vec{a},s,s')$ means that the system may reach situation $s'$
when the agent executes the sequence of agent actions $\vec{a}$
depending on environment reactions:

\noindent
\begin{small}
\[\begin{array}{l} 
    \hspace{-0.5em}
    Do_{ag}(\epsilon,s,s') \doteq s'=s    \mbox{ (where $\epsilon$ is the empty sequence of actions)}
    \\[1ex]
		\hspace{-0.5em}
    Do_{ag}([A(\vec{x}),\sigma],s,s') \doteq\\
    \quad\exists e. Poss(A(\vec{x},e),s) \land Do_{ag}(\sigma,do(A(\vec{x},e),s),s')
\end{array} \]
\end{small}%

%

A \emph{strong plan} for the agent guarantees the achievement of a goal no matter
how the environment reacts; it is a strategy for the agent to follow
to ensure that the goal is achieved. (DL21) define a strategy as a
function from situations to (instantiated) agent actions. That is,
$\mathit{f}(s) = A(\vec{t})$ denotes that the strategy $\mathit{f}$ applied to situation
$s$ returns $A(\vec{t})$
as the next action to do.
The special agent action  $stop$ (with no effects and preconditions) may be returned when the strategy
wishes to stop.
Given a strategy, we can check whether it forces the goal to become true in spite of the
environment reactions, i.e., is a strong plan to achieve the goal.
Formally, we have
$AgtCanForceBy(Goal, s, {f})$, i.e., the agent can force $Goal$ to
become true by following strategy $\mathit{f}$ in $s$:


\noindent
\begin{small}
\[\begin{array}{l} 
AgtCanForceBy(Goal,s,\mathit{f}) \doteq \forall P.[\ldots  \limp P(s)] \\
\mbox{ where \ldots stands for } \\
\lbrack (\mathit{f}(s)=stop \land  Goal(s)) \limp P(s)  \rbrack \land \\
\lbrack \exists A. \exists \vec{t}.(\mathit{f}(s)=A(\vec{t}) \neq stop \land  Poss_{ag}(A(\vec{t}),s)  \land \\
\hspace*{1.5em}   \forall e.(Poss (A(\vec{t},e),s) \limp P(do(A(\vec{t},e),s)))) \\
\limp P(s) \rbrack 
\end{array} \]
\end{small}%

\noindent
We say that $AgtCanForce(Goal,s)$ holds iff there is a strategy $\mathit{f}$ such that $AgtCanForceBy(Goal,s,\mathit{f})$ holds.

\paragraph{Example Cont.} 
Strategies use the $stop$ action, hence we define a $stop_{LL}$ action
at the low level;
this action always terminates with the
$Success_{LS}$ reaction.
The strategy that guarantees reaching location $13$ is defined as follows:


\vspace{-0.55em}
\noindent
\begin{small}
\[\begin{array}{l}
\hspace{-0.5em}
\D_l^{tt} \models AgtCanForceBy(At_{LL}(13),S_0,\mathit{f_l}) \\
\hspace{-0.5em} \mbox{where } \\
\hspace{-0.5em}
\mathit{f_l}(s) \doteq
    \begin{cases}
      stop_{LL} & \hspace{-0.45em} \text{if  } At_{LL}(13,s) \\
      \mathit{fixFlatTire(l)} & \hspace{-0.45em} \text{if  } At_{LL}(l,s) \land l \neq 13 \land \mathit{Flat_{LL}}(s)\\
			drive(o,d) & \hspace{-0.45em} \text{if  }  At_{LL}(o,s) \land o \neq 13 \land \neg \mathit{Flat_{LL}}(s)  \\
		  &	\hspace{-0.45em} \land \; Spare_{LL}(d,s) \land Road_{LL}(o,d,s)  \\
			& \hspace{-0.45em} \land \neg \mathit{Visited_{LL}}(d) \\
			wait_{LL} & \hspace{-0.45em} \text{otherwise}
    \end{cases}       
\end{array} \]
\end{small}%

\noindent
That is, the agent should stop if she is already at location $13$,
otherwise she drives to a location she has not visited before and that
has a spare tire; in case of a flat tire, action $\mathit{fixFlatTire}$ is
executed; in all other cases the agent waits.

\paragraph{High-Level Programs and \ConGolog.}
To represent and reason about complex actions or processes obtained by suitably
executing atomic actions, various so-called \emph{high-level programming
languages} have been defined.
%
%
Here we concentrate on (a variant of) \ConGolog
\cite{DBLP:journals/ai/GiacomoLL00} that includes the following constructs: 

\begin{small}
 \[ \delta ::= \alpha  \mid  \varphi?  \mid  \delta_1;\delta_2  \mid
  \delta_1 \ndet \delta_2  \mid  \pi x.\delta  \mid  \delta^*  \mid 
  \delta_1 \conc \delta_2 \]
\end{small}%

\noindent
In the above, $\alpha$ is an action term, possibly with parameters,
and  $\varphi$ is a 
situation-suppressed formula, i.e., a formula 
with all situation arguments in fluents suppressed. 
As usual, we denote by $\varphi[s]$ the 
formula obtained from $\varphi$ by restoring the situation
argument $s$ into all fluents in $\varphi$.
The sequence of program $\delta_1$ followed by program $\delta_2$ is denoted by $\delta_1;\delta_2$.
Program $\delta_1 \ndet \delta_2$ allows for the nondeterministic choice
between programs $\delta_1$ and $\delta_2$, while $\pi x.\delta$ executes program
$\delta$ for \textit{some} nondeterministic choice of a  
binding for object variable $x$ (observe that such a choice is, in general, unbounded).  $\delta^*$
performs $\delta$ zero or more times.
Program $\delta_1 \conc \delta_2$ expresses the concurrent execution
(interpreted as interleaving) of programs $\delta_1$ and $\delta_2$.
%
The construct \textbf{if} $\phi$ \textbf{then} $\delta_1$
\textbf{else} $\delta_2$ \textbf{endIf} is defined as $[\phi?;
\delta_1 ] \mid [\neg \phi?; \delta_2]$. 
We also use $nil$, an abbreviation for $True?$, to represents the
\emph{empty program}, i.e., when nothing remains to be performed.

Formally, the semantics of \ConGolog\ is specified in terms of
single-step transitions, using the following two predicates:
\emph{(i)} $Trans(\delta,s,\delta',s')$, which holds if one step
of program $\delta$ in situation $s$ may lead to situation $s'$ with
$\delta'$ remaining to be executed; and \emph{(ii)}  $Final(\delta,s)$, which
holds if program $\delta$ may legally terminate in situation $s$.
The definitions of $Trans$ and $Final$ we use are as in
\cite{DBLP:conf/kr/GiacomoLP10}; 
differently from \cite{DBLP:journals/ai/GiacomoLL00}, 
the test construct $\varphi?$ does not yield
any transition, but is final when satisfied
(see the appendix for details).

Predicate $Do(\delta,s,s')$ means that program $\delta$, when executed starting in situation $s$, has as a legal terminating situation $s'$, and is defined as $Do(\delta,s,s') \doteq \exists \delta'. Trans^*(\delta,s,\delta',s') \land Final(\delta',s')$ where $\Trans^*$ denotes the reflexive transitive closure of $\Trans$.
We use  $\CGaxioms$ to denote the axioms defining the \ConGolog\
programming language.

For simplicity in this paper, 
we use a restricted class of \ConGolog programs which are \emph{situation-determined} (SD) \cite{DBLP:conf/aamas/GiacomoLM12}, i.e., for every
sequence of actions, the remaining program is uniquely
determined by the resulting situation:


\noindent
\begin{small}
$\SitDet(\delta,s) \doteq \forall s',\delta', \delta''. \\$
$\hspace*{3em}\Trans^*(\delta,s, \delta',s') \land \Trans^*(\delta,s, \delta'',s') \limp$
$\delta'=\delta''$
\end{small}%
%
%

%

\paragraph{\ConGolog Program Execution in ND Domains.}
To specify agent behaviors in nondeterministic domains, (DL21) use
\ConGolog\ \emph{agent programs}, which are composed as usual, but only
involve agent (reaction-suppressed) atomic actions.
The semantics for such agent programs is a variant of the one above
where a transition for an atomic agent 
action may occur whenever there exists a reaction that can produce
it (see appendix for details.)

\paragraph{Strategic Reasoning in Executing Programs in ND Domains.}
To represent the ability of the agent to execute an agent program in a
ND domain (DL21) introduce $AgtCanForceBy(\delta, s, \mathit{f})$ as an
adversarial version of $Do$ in presence of environment reactions. This
predicate states that strategy $\mathit{f}$, a function from situations to
agent actions (including the special action $stop$), executes 
SD \ConGolog\ agent program 
$\delta$ in situation $s$ considering its
nondeterminism angelic, as in the standard $Do$, but also considering
the nondeterminism of environment reactions devilish/adversarial:

\noindent
\begin{small}
\[\begin{array}{l} 
AgtCanForceBy(\delta, s, \mathit{f}) \doteq \forall P.[\ldots  \limp P(\delta, s)] \\
\mbox{ where \ldots stands for } \\
\lbrack (\mathit{f}(s)=stop \land  Final(\delta,s)) \limp P(\delta, s)  \rbrack \land {}\\
\lbrack \exists A. \exists \vec{t}.(\mathit{f}(s)=A(\vec{t}) \neq stop \land  {}\\
\exists e. \exists \delta'. Trans(\delta,s,\delta',do(A(\vec{t},e),s))  \land {}\\
\forall e.(\exists \delta'. Trans(\delta,s,\delta',do(A(\vec{t},e),s))) \limp \\
\exists \delta'. Trans(\delta,s,\delta',do(A(\vec{t},e),s)) \land P(\delta',do(A(\vec{t},e),s)) \\
\limp P(\delta,s) \rbrack 
\end{array} \]
\end{small}%

\noindent
We say that predicate $AgtCanForce(\delta, s)$ holds iff there exists
a strategy $\mathit{f}$ s.t.\ $AgtCanForceBy(\delta, s, \mathit{f})$ holds.

\paragraph{Example Cont.} Suppose we have a program $\delta_{go}^{LL}(l)$ which lets the agent drive to any location, or fix a flat any number of times until a destination $l$ is reached:
\begin{small} $\delta_{go}^{LL}(l) = (\pi o,d. drive(o,d) \mid \pi d. \mathit{fixFlatTire}(d))^*; At_{LL}(l)?$ \end{small}%

Now suppose the agent has been assigned the task of going to location
$13$. We can show that she has a strategy for executing the task $\delta_{go}^{LL}(13)$:  

\noindent
\begin{small}
\[\begin{array}{l} 
\D^{tt}_l \models AgtCanForceBy(\delta_{go}^{LL}(13),S_0,\mathit{f})
\end{array} \]
\end{small}%

\noindent 
Here, the strategy $\mathit{f}$ to execute the task is same as $\mathit{f_l}$ above. 

%


%
\section{Abstraction in Nondeterministic Domains} \label{sec:NDabstraction} 



In this section, we show how we can extend the agent abstraction framework of
(BDL17) to handle nondeterministic domains.
%
As in (BDL17), we assume that
there is a high-level/abstract (HL) action
theory $\D_h$ and a low-level/concrete (LL) action theory 
$\D_l$ representing the agent's possible behaviors at different levels
of detail.  In (BDL17), these are standard BATs; here, we assume
that they are both NDBATs.
$\mathit{\D_h}$ (resp. $\mathit{\D_l}$) involves a finite set of primitive action types
$\mathit{\A_h}$ (resp. $\mathit{\A_l}$) and a finite set of primitive
fluent predicates $\mathit{\F_h}$ (resp. $\mathit{\F_l}$).
The terms of sort \emph{Object} are assumed be a
countably infinite set $\N$ of standard names for which we have the unique
name assumption and domain closure.
We also assume that \emph{Reaction} is a sub-sort of \emph{Object}.
Also, $\mathit{\D_h}$ and $\mathit{\D_l}$ are assumed to share no
domain specific symbols except for the set of standard names for
objects $\mathit{\N}$.
For simplicity and w.l.o.g., it is assumed that there are no functions
other than constants and no non-fluent predicates.

\paragraph{Refinement Mapping.}
As in (BDL17), we assume that one relates the HL and LL theories by
defining a \emph{refinement mapping} that specifies how HL atomic
actions are implemented at the low level and how HL fluents can be
translated into LL state formulas.
In deterministic domains, one can simply map HL atomic actions to a LL
program that the agent uses to implement the action.
In nondeterministic domains however, we additionally need to specify 
what HL reaction the environment performs in each LL refinement of the HL action.
%
\begin{definition} [NDBAT Refinement Mapping]
A \emph{NDBAT refinement mapping} $m$ is a triple
$\langle m_a, m_s, m_f \rangle$ where $m_a$
associates each
HL primitive action type $\mathit{A}$ in
$\mathit{\A_h}$ to a
SD \ConGolog \emph{agent program} $\delta^{ag}_A$ defined over the
LL theory 
that implements the agent (i.e., reaction-suppressed) action, i.e., $m_a(A(\vec{x})) =
\delta^{ag}_A(\vec{x})$,
$m_s$ associates each $\mathit{A}$ to a
SD \ConGolog \emph{system program}
 $\delta^{sys}_A$ defined over the
LL theory 
that implements the system action, i.e., $m_s(A(\vec{x},e)) =
\delta^{sys}_A(\vec{x},e)$, thus specifying when the HL reaction
occurs (system programs are interpreted using the standard transition semantics),
and (as in (BDL17)) $m_f$ 
maps each situation-suppressed HL fluent
$\mathit{F}(\vec{x})$  in $\mathit{\F_h}$ to a situation-suppressed formula
$\phi_F(\vec{x})$ defined over the LL theory that characterizes
the concrete conditions under which $\mathit{F}(\vec{x})$ holds in a situation,
i.e., $m_f(\vec{x}) = \phi_F(\vec{x})$.
\end{definition}

We can extend a mapping to a sequence of agent actions in the
obvious way, i.e.,  $m_a(\alpha_1, \ldots, \alpha_n) \doteq m_a(\alpha_1);\ldots;
m_a(\alpha_n)$ for $n \geq 1$ and  $m_a(\epsilon) \doteq nil$, and
similarly for sequences of system actions.
Note that $m_a(\vec{\alpha})$ is well defined even if the action parameters are
not ground; but the action functions must be given.
We also extend the notation so that $m_f(\phi)$ stands for the result
of substituting every fluent $\mathit{F}(\vec{x})$ in situation-suppressed formula $\phi$ by $m_f(\mathit{F}(\vec{x}))$.

Agent actions and system actions must be mapped in a consistent way.
To ensure this, we require the following:

\begin{constraint} \label{cstr:ProperMapp}
(Proper Refinement Mapping)\\
For every high-level system action
sequence $\vec{\alpha}$ and every high-level action $A$, we have that:

\vspace{-0.65em}
\begin{small}  
\[\begin{array}{l} 
  \D_l  \cup \C \models \forall s. (Do(m_s(\vec{\alpha}),S_0,s) \limp\\
\forall \vec{x},s'. (Do_{ag}(m_a(A(\vec{x})),s,s') \equiv \exists e. Do(m_s(A(\vec{x},e)),s,s')))
  \end{array}\]
\end{small}%

\end{constraint}

\noindent
This ensures that (1) for every situation $s'$ that can be reached
from $s$ by executing a refinement of the HL agent action
$A(\vec{x})$, there is a HL reaction $e$ that generates it,
and (2) that for every situation $s'$ that can be reached from
$s$ by executing a refinement of the HL system action
$A(\vec{x},e)$, there is a refinement of the HL agent action
that generates it
(in the above, $s$ is any situation reached by executing a sequence of
HL system actions).
Note that (1) essentially states that the reaction existence
requirement holds at the low level for refinements of HL actions and
(2) amounts to having the reaction independence requirement hold.
We say that an NDBAT refinement mapping $m$ is \emph{proper wrt low-level
NDBAT} $\D_l$ if this constraint holds.


\paragraph{Example Cont.} NDBAT $\D_h^{tt}$
abstracts over driving and fixing potential flat tires.  Action
$\mathit{driveAndTryFix}(o,d,r)$ can be performed to drive from origin 
$o$ to destination  $d$, and fix a flat if one occurs and a
spare is available. This action is executable when
$\mathit{driveAndTryFix}(o,d)$ is executable, i.e., when the agent is at $o$, a
road connects $o$ to $d$, and agent does not have a flat tire;
the environment reaction $r$ may take on the following values:
$\mathit{DrvNoFlat}$, if the tire does not go flat, $\mathit{DrvFlat}$ if the tire goes
flat and it is not possible to fix it, or $\mathit{DrvFlatFix}$ if the tire
went flat and the flat was fixed. Action $wait_{HL}$ lets the agent remain idle at a location.
$\D_h^{tt}$ includes the following system and agent action precondition axioms:

\noindent
\begin{small}
\[\begin{array}{l} 
Poss_{ag}(\mathit{driveAndTryFix}(o, d), s) \doteq  \\
\hspace*{1.3em} o \neq d \land At_{HL}(o,s) \land Road_{HL}(o,d,s) \land \neg \mathit{Flat_{HL}}(s)
\\
Poss_{ag}(wait_{HL}, s) \doteq TRUE
\\
Poss(driveAndTryFix(o,d,r), s) \equiv \\
\hspace*{1.3em} Poss_{ag}(\mathit{driveAndTryFix}(o,d), s) \land {}\\
\hspace*{1.3em} (r=\mathit{DrvNoFlat} \lor {}\\
\hspace*{1.8em} \neg Spare_{HL}(d,s) \land r=\mathit{DrvFlat} \lor {}\\
\hspace*{1.8em} Spare_{HL}(d,s) \land r=\mathit{DrvFlatFix})
\\
Poss(wait_{HL}(r), s) \equiv  \\
\hspace*{1.3em} Poss_{ag}(wait_{HL}, s) \land r=Success_{HW}
\end{array} \]
\end{small}%

$\D_h^{tt}$ also includes the following SSAs:

\noindent
\begin{small}
\[\begin{array}{l}
At_{HL}(l, do(a,s)) \equiv 
\exists o,r. a = \mathit{driveAndTryFix}(o, l, r) \; \lor \\
\hspace*{1.3em}
At_{HL}(l, s) \; \land \forall d,r. a \neq  \mathit{driveAndTryFix}(l, d, r)
\\
\mathit{Flat_{HL}}(do(a,s)) \equiv \\
\hspace*{1.3em}
\exists o, d. a=\mathit{driveAndTryFix}(o,d, \mathit{DrvFlat}) \; \lor 
\mathit{Flat_{HL}}(s) 
\\
\mathit{Visited_{HL}}(l,do(a,s)) \equiv \\
\hspace*{1.3em} 
\exists o,r.  a=\mathit{driveAndTryFix}(o,l,r) \; \lor \mathit{Visited_{HL}}(l,s)
\end{array} \]
\end{small}%

\noindent
For the other fluents, we have SSAs specifying that they are
unaffected by any action.

The initial state axioms of $\D_h^{tt}$ are same as $\D_l^{tt}$, with high-level fluents
having a distinct name (with HL suffix); e.g., $Road_{HL}$ is axiomatized exactly as $Road_{LL}$.

We specify the relationship between the HL and LL NDBATs through the following refinement mapping $m^{tt}$:

\begin{small}
\[\begin{array}{l}
m^{tt}_a(\mathit{driveAndTryFix}(o,d)) = \\
\quad drive(o, d); \\
\quad \mbox{\textbf{if} } \neg \mathit{Flat}_{LL} \mbox{ \textbf{then} }  nil 
\mbox{\textbf{ else}} \\
\quad \hspace*{1em} \mbox{\textbf{if}} \neg Spare_{LL}(d) \mbox{ \textbf{then} }  nil  
\mbox{\textbf{ else} } \mathit{fixFlatTire}(d) 
\mbox{\textbf{ endIf}} \\
\quad \mbox{\textbf{endIf}}
\\
m^{tt}_a(wait_{HL})=wait_{LL} \\[1ex]
m^{tt}_s(\mathit{driveAndTryFix}(o,d,r_h)) = \\
\quad \pi r_l.drive(o, d, r_l); \\
\quad \mbox{\textbf{if} } \neg \mathit{Flat_{LL}} \mbox{ \textbf{then}
    } r_h= DrvNoFlat?
\mbox{\textbf{ else}} \\ 
\quad \hspace*{1em} \mbox{\textbf{if}} \neg Spare_{LL}(d) \mbox{ \textbf{then} } r_h= \mathit{DrvFlat}? \\
\quad \hspace*{1em} \mbox{\textbf{else} } \mathit{fixFlatTire}(d, Success_{LF}); r_h= \mathit{DrvFlatFix}? \\
\quad \hspace*{1em}\mbox{\textbf{endIf}} \\
\quad \mbox{\textbf{endIf}}
\\
m^{tt}_s(wait_{HL}(r_h))=\pi r_l.wait_{LL}(r_l);r_h=Success_{HW}?\\[1ex]
m^{tt}_f(\mathit{Flat}_{HL})=\mathit{Flat}_{LL} 
\end{array} \]
\end{small}%

\noindent
Thus, the HL agent action $driveAndTryFix(o,d)$ is implemented by an LL
program where the agent first performs $drive(o,d)$, and depending on
whether the tire has gone flat and a spare exists at location $d$, fixes the tire or does nothing. Fluents $At_{HL}(l)$, $Spare_{HL}(l)$, $Road_{HL}(o,d)$, $Dest_{HL}(l)$, and $\mathit{Visited_{HL}}(l)$ are mapped to their low-level counterparts similar to $\mathit{Flat}_{HL}$.
%
%
%
We can show that:\footnote{For proofs, see appendix.}
\begin{proposition} \label{prop:ProperRefEx}
NDBAT refinement mapping $m^{tt}$ is \emph{proper} wrt $\D_l^{tt}$.
\end{proposition}

\paragraph{$m$-Bisimulation.} To relate the HL and LL
models/theories, (BDL17) define a variant of bisimulation
\cite{DBLP:conf/ijcai/Milner71,DBLP:books/daglib/0067019}.\footnote{As usual, $M,v \models \phi$ means that model $M$ and assignment $v$
satisfy  formula $\phi$ (where  $\phi$ may contain free variables that are interpreted by $v$); also $v[x/e]$ represents the variable
assignment that is just like $v$ but assigns variable $x$ to entity $e$.}
The base condition for the bisimulation is:
\begin{definition} [$m$-isomorphic situations]
Let $M_h$ be model of the HL theory $\D_h$, and $M_l$ a model of the LL theory $\D_l \cup \C$.
We say that situation $s_h$ in $M_h$ is $m$-\emph{isomorphic} to situation
$s_l$ in $M_l$, written $s_h \sim_m^{M_h,M_l} s_l$, if and only if
\[M_h,v[s/s_h] \models F(\vec{x},s) \mbox{ iff } M_l,v[s/s_l] \models
  m(F(\vec{x}))[s]\]
for every high-level primitive fluent
$F(\vec{x})$ in $\mathit{\F_h}$ and every variable assignment $v$.
  \end{definition}

\noindent
If $s_h \sim_m^{M_h,M_l} s_l$, then $s_h$ and $s_l$ evaluate all
HL fluents the same.

\begin{definition} [$m$-Bisimulation]
Given $M_h$ a model of $\D_h$, and $M_l$ a model of $\D_l \cup \C$,
a relation $B \subseteq \Delta_S^{M_h} \times \Delta_S^{M_l}$ (where $\Delta_S^{M}$ stands for the situation domain of $M$)  is an
\emph{$m$-bisimulation relation between $M_h$ and $M_l$}  if $\langle
s_h, s_l \rangle \in B$ implies that:
$(i)$ $s_h \sim_m^{M_h,M_l} s_l$;
$(ii)$  for every HL primitive action type $\mathit{A}$ in $\mathit{\A_h}$, if  there exists $s_h'$ such that $M_h,v[s/s_h,s'/s_h'] \models  Poss(A(\vec{x}),s) \land s' = do(A(\vec{x}),s)$, 
then there exists  $s_l'$ such that $M_l,v[s/s_l,s'/s_l'] \models
Do(m(A(\vec{x})),s,s')$ and $\langle s_h',s_l' \rangle \in B$; and
$(iii)$ for every HL primitive action type $\mathit{A}$ in  $\mathit{\A_h}$, if there exists  $s_l'$ such that 
  $M_l,v[s/s_l,s'/s_l'] \models Do(m(A(\vec{x})),s,s')$, then there exists $s_h'$ such that
  $M_h,v[s/s_h,s'/s_h'] \models Poss(A(\vec{x}),s) \land s' =   do(A(\vec{x}),s)$ and $\langle s_h',s_l' \rangle \in B$. 
$M_h$ is \emph{$m$-bisimilar} to $M_l$, written $M_h \sim_m M_l$, if and only if there exists an
$m$-bisimulation relation $B$ between $M_h$ and $M_l$ such that
$(S_0^{M_h}, S_0^{M_l}) \in B$.
\end{definition}

The definition of $m$-bisimulation can remain as in (BDL17)
where conditions (ii) and (iii) are applied
to high-level \emph{system} primitive actions and their mapping $m_s$.

The definition of $m$-bisimulation ensures that performing a
HL system action results in $m$-bisimilar situations.
We can show that with the restriction to proper mappings, this
automatically carries over to HL agent actions (so there is
no need to change the definition of $m$-bisimulation to get this):

\begin{theorem} \label{thm:bisimProperMgeneralizes2agtAct} 
  Suppose that  $M_h \sim_m M_l$, where  $M_h \models \D_h$, $M_l
  \models \D_l \cup \C$, $\D_h$ and $\D_l$ are both NDBATs, and $m$ is
  proper wrt $D_l$.
  Then for any HL system action sequence $\vec{\alpha}$ and any HL
      primitive action $A$, we have that:
  \begin{enumerate}
  \item
      if there exist  $s_h$ and $s_h'$ such that
      $M_h,v[s/s_h,s'/s_h'] \models Do(\vec{\alpha},S_0,s) \land
      Do_{ag}(A(\vec{x}),s,s')$,
      then there exist  $s_l$ and $s_l'$ such that $M_l,v[s/s_l,s'/s_l'] \models
      Do(m_s(\vec{\alpha}),S_0,s) \land 
      Do_{ag}(m_a(A(\vec{x})),s,s')$, 
$s_h \sim_m^{M_h,M_l} s_l$, and $s_h' \sim_m^{M_h,M_l} s_l'$;

    \item
      if there exist $s_l$ and $s_l'$ such that $M_l,v[s/s_l,s'/s_l'] \models
      Do(m_s(\vec{\alpha}),S_0,s) \land  Do_{ag}(m_a(A(\vec{x})),s,s')$,
      then there exist $s_h$ and $s_h'$ such that
      $M_h,v[s/s_h,s'/s_h'] \models  Do(\vec{\alpha},S_0,s) \land
      Do_{ag}(A(\vec{x}),s,s')$, 
$s_h \sim_m^{M_h,M_l} s_l$, and $s_h' \sim_m^{M_h,M_l} s_l'$.
    \end{enumerate}
\end{theorem}

(BDL17) use $m$-bisimulation to define notions of sound/complete
abstraction between a high-level action theory and a low-level one: $\D_h$ is a \emph{sound abstraction of} $\D_l$
\emph{relative to refinement mapping} $m$ if and only if, for all models $M_l$ of $\D_l \cup \C$, there exists a model $M_h$ of $\D_h$ such that  $M_h \sim_m M_l$. With a sound abstraction,
whenever the high-level theory entails that a sequence of actions
is executable and achieves a certain condition, then the
low level must also entail that there exists an executable refinement
of the sequence such that the ``translated'' condition
holds afterwards. Moreover, whenever the low level considers
the executability of a refinement of a high-level action is
satisfiable, then the high level does also. A dual notion is also defined: $\D_h$ is a \emph{complete abstraction of} $\D_l$ \emph{relative to refinement mapping} $m$ if and only if, for all models $M_h$ of $\D_h$, there exists a model $M_l$ of $\D_l \cup \C$ such that  $M_l \sim_m M_h$.

The notion of $m$-bisimulation provides the semantic underpinning of
these notions of abstraction.  (BDL17) also prove the following results that
identify a set of properties that are necessary and sufficient to have
a sound/complete abstraction (here we adjust the notation to use system
actions) and can be used to verify that one has a sound/complete abstraction,
e.g., by model checking or theorem proving techniques:


\begin{theorem}[BDL17] \label{thm:verifySound}  
 $\D^h$ is a sound abstraction of $\D^l$ relative to mapping $m$ if and only if for any sequence of high-level system actions $\vec{\alpha}$: 

\begin{small}
\begin{description}
\item[(a)]
$\D^l_{S_0} \cup \D^l_{ca} \cup \D^l_{coa} \models m_f(\phi)$, for all
$\phi \in D^h_{S_0}$,
\item[(b)]
$\mathit{\D^l \cup \C \models \forall s. Do(m_s(\vec{\alpha}),S_0,s) \limp}$ \\
\hspace*{0.2em} $\bigwedge_{A_i \in \A^h} \forall \vec{x},r_h.
 (m_f(\phi^{Poss}_{A_i}(\vec{x},r_h))[s] 
    \equiv  \\
\hspace*{0.3em} \exists s' Do(m_s(A_i(\vec{x},r_h)),s,s')),$
\item[(c)]
$\D^l \cup \C \models \forall s. Do(\vec{\alpha},S_0,s) \limp $ \\
\hspace*{0.2em} $\bigwedge_{A_i \in \A^h}  \forall \vec{x},r_h, s'. (Do(m_s(A_i(\vec{x},r_h)),s,s') \limp$ \\
\hspace*{1em} $\bigwedge_{F_i \in \F^h}  \forall \vec{y}  (m_f(\phi^{ssa}_{F_i,A_i}(\vec{y},\vec{x},r_h))[s] \equiv m_f(F_i(\vec{y}))[s'])),$
\end{description}
\end{small}%
\noindent
where $\phi^{Poss}_{A_i}(\vec{x},r_h)$ is the right hand side (RHS) of the precondition axiom for system action $A_i(\vec{x},r_h)$, and $\phi^{ssa}_{F_i,A_i}(\vec{y},\vec{x},r_h)$ is the RHS of the successor state axiom for $F_i$ instantiated with system action
$A_i(\vec{x},r_h)$ where action terms have been eliminated using $\D^h_{ca}$. 
\end{theorem}



\begin{theorem}[BDL17]\label{thm:verifySoundComplete}
If $\D^h$ is a sound abstraction of $\D^l$ relative to mapping $m$, then
$\D^h$ is also a complete abstraction of $\D^l$ wrt mapping $m$
if and only if for every model $M_h$ of $\D^h_{S_0} \cup \D^h_{ca} \cup \D^h_{coa}$,
there exists a model  $M_l$ of $\D^l_{S_0} \cup \D^l_{ca}
\cup\D^l_{coa}$ such that $S_0^{M_h} \sim_m^{M_h,M_l} S_0^{M_l}$. 
\end{theorem} 


%


Using the above results, we can show that: 

\begin{proposition} \label{prop:SoundCompEx}
$\D^{tt}_h$ is a sound and complete abstraction of $\D^{tt}_l$ wrt $m^{tt}$.
\end{proposition}

Note that in this paper, we are focusing on fully observable domains, so we
will just present results about $m$-bisimilar NDBAT models;
$M_h \sim_m M_l$ essentially means that $M_h$ is a sound and complete abstraction of $M_l$
relative to $m$.



%
\section{Results about Action Executions} \label{sec:ResultsActExec}

We now show some interesting results about the use of NDBAT
abstractions to reason about action executions.
Our results
will be mostly about $m$-bisimilar NDBAT models, i.e., where
$M_h \sim_m M_l$;
unless stated otherwise, we assume we have NDBATs
$\D_h$ and $\D_l$, that $M_h \models \D_h \cup \C$ and $M_l \models \D_l \cup \C$, and
that $m$ is proper wrt $\D_l$.

%
Firstly, we have that
$m$-isomorphic situations satisfy the same high-level situation-suppressed formulas:
\begin{lemma}[BDL17] \label{lem:sitSupBisim} 
  If $s_h \sim_m^{M_h,M_l} s_l$, then for any high-level
  situation-suppressed formula $\phi$, we have that:
\\[0.5ex]  
\begin{small}  
\hspace*{1em} $M_h,v[s/s_h] \models \phi[s]\ \ \mbox{if and only if}\ \ 
M_l,v[s/s_l] \models m_f(\phi)[s].$
\end{small}%
\end{lemma}

Secondly, we can show that in $m$-bisimilar
models, the same sequences of high-level system actions are executable, and
that in the resulting situations, the same high-level
situation-suppressed formulas hold:
\begin{theorem} \label{thm:bisimL2HOffND}
  If  $M_h \sim_m M_l$, then 
  for any sequence of high-level system actions $\vec{\alpha}$
  and any high-level situation-suppressed formula $\phi$, 
  we have that

\begin{small}  
\[\begin{array}{l}
  M_l,v \models \exists s'. Do(m_s(\vec{\alpha}),S_0,s') \land
  m_f(\phi)[s'] \hspace{1em} \mbox{ if and only if }\\
  \hspace{2em} M_h,v \models Executable(do(\vec{\alpha},S_0)) \land
  \phi[do(\vec{\alpha},S_0)].
\end{array}\]
\end{small}%
\end{theorem}

\noindent
Here and below, sequences of high-level system actions $\vec{\alpha}$
may contain free variables in the action parameters, but the action
functions must be given, and similarly for sequences of high-level
agent actions (this generalizes (BDL17) which only considers
ground action sequences).

We can extend this result to sequences of \emph{agent actions} by exploiting
the fact that $m$ is a proper mapping:
\begin{theorem} \label{thm:bisimL2HAgtActOffND} 
  If  $M_h \sim_m M_l$, then 
  for any sequence of high-level agent actions $\vec{\alpha}$ and
  any high-level situation-suppressed formula $\phi$, 
  we have that

\begin{small}  
\[\begin{array}{l}
  M_l,v \models \exists s'. Do_{ag}(m_a(\vec{\alpha}),S_0,s') \land
  m_f(\phi)[s'] \hspace{1em} \mbox{ if and only if }\\
  \hspace{2em} M_h,v \models \exists s'. Do_{ag}(\vec{\alpha},S_0,s') \land
  \phi[s'].
\end{array}\]
\end{small}%
\end{theorem}

\noindent
This means that if the agent has a weak plan to achieve a goal at the high
level, then there exists a refinement of it that is a weak plan to
achieve the mapped goal at the low level.

\paragraph{Example Cont.} At the high level, a weak plan to achieve
the goal of
getting to location 13 
is to first drive to $12$ and then to $13$: 

\begin{small}  
  \[\begin{array}{l}
      \D_h^{tt} \models \exists s'. Do_{ag}(\vec{\alpha},S_0,s') \land
      At_{HL}(13,s') \mbox{   where } \\
			\vec{\alpha}= \mathit{driveAndTryFix}(11,12);\mathit{driveAndTryFix}(12,13)
  \end{array}\]
\end{small}%

\noindent
This plan works provided no flat occurs when driving to $12$, as
there is no spare there.
By Th.~\ref{thm:bisimL2HAgtActOffND}, there exists a weak plan at the
low level that refines this high-level plan: 

\begin{small}  
  \[\begin{array}{l}
			\D^{tt}_l \models \exists
      s'. Do_{ag}(m_a(\vec{\alpha}),S_0,s') \land  At_{LL}(13,s')
      \land {}\\
     \hspace{4.7em} Do_{ag}([drive(11,12);drive(12,13)],S_0,s') 
  \end{array}\]
\end{small}%

\section{Results about Strategic Reasoning}\label{sec:ResultsFOND}

Let us now discuss how NDBAT abstractions can be used in FOND domains
to synthesize strategies to fulfill reachability/achievement goals as well as
temporally extended goals/tasks.
%
First of all, we need to consider how much strategic
reasoning the agent needs to do to execute a high-level
atomic action at the low level.
A given HL agent action $A(\vec{x})$ is
mapped to a LL agent program $m_a(A(\vec{x}))$ that implements it.
As we have seen, in $m$-bisimilar models, Constraint \ref{cstr:ProperMapp} ensures that
if $A(\vec{x})$ is executable at the HL, then there exists a terminating execution of
$m_a(A(\vec{x}))$ at the LL
(this holds for HL system actions as well).
But this does not mean that all executions of $m_a(A(\vec{x}))$
terminate, as some may block or diverge, due to either agent or environment choices.
%
In general, the agent must do strategic reasoning to ensure that the
execution of $m_a(A(\vec{x}))$ terminates (and the environment may need
to cooperate as well).  But we can impose further constraints on the
mapping of HL actions to avoid this or ensure that a strategy exists.
Note that ensuring that the
execution of $m_a(A(\vec{x}))$ terminates does not mean that the agent
controls the action's outcome (e.g., not having a flat); the outcome is still determined by the environment reactions.

\paragraph{Constraints on HL action implementation.}
First, we may want to require that the mapping of HL actions is such
that the implementation program always terminates and no LL strategic
reasoning is required to ensure termination.
To do this, we first define:





\begin{small}
\[\begin{array}{l}
InevTerminates(\delta, s) \doteq \forall P.[\ldots  \limp P(\delta, s)] \\
\mbox{ where \ldots stands for } \\
\lbrack (Final(\delta,s)) \limp P(\delta, s)  \rbrack \land \\
\lbrack \exists s'. \exists \delta'. Trans(\delta,s,\delta',s')  \land \\
\forall s'.\forall \delta'. (Trans(\delta,s,\delta',s') \limp P(\delta',s')) \\
\limp P(\delta,s) \rbrack
\end{array} \]
\end{small}%

\noindent
$InevTerminates(\delta,s)$ means that program $\delta$ executed
starting in situation $s$ \emph{inevitably terminates} , i.e., all its
executions eventually terminate. \footnote{This is analogous to the CTL
\cite{DBLP:conf/lop/ClarkeE81} formula $AF\phi$, i.e., on all paths eventually $\phi$. } 

Then we can ensure that for any HL agent action that is possibly executable at the low
level, all executions terminate (i.e., they never block or diverge) by
requiring the following:

\begin{constraint} \label{cstr:HLactionsNecTerminate}
(HL actions Inevitably Terminate)\\
For every high-level system action
sequence $\vec{\alpha}$ and every high-level action $A$, we have that:

\begin{small}
  \[\begin{array}{l}
  \D_l \cup \C \models \forall s. (Do(m_s(\vec{\alpha}),S_0,s) \limp \forall \vec{x}.\\
(\exists s'.Do_{ag}(m_a(A(\vec{x})),s,s') \supset InevTerminates(m_a(A(\vec{x})),s)))
  \end{array}\]
\end{small}%
\end{constraint}

\noindent
\begin{proposition} \label{prop:NecTerm}
NDBAT $\D_l^{tt} $ and mapping $m^{tt}$ satisfy  Constraint \ref{cstr:HLactionsNecTerminate}.
\end{proposition}

If we impose Constraint \ref{cstr:HLactionsNecTerminate}, then the agent
that executes the program that implements the HL action can be a dumb executor.
But this may seem too restrictive.  An alternative is to 
impose the weaker requirement that
for any HL agent action that is possibly executable at the LL,
the agent has a strategy to execute it to termination no
matter how the environment reacts (even if not controlling its
outcome).  Formally:

\begin{constraint} \label{cstr:AgtAlwsCanExecuteHLactions}
(Agt Can Always Execute HL actions)\\
For every high-level system action
sequence $\vec{\alpha}$ and every high-level action $A$, we have that:

\begin{small}
  \[\begin{array}{l}
  \D_l \cup \C \models \forall s. (Do(m_s(\vec{\alpha}),S_0,s) \limp \forall \vec{x}.\\
(\exists s'.Do_{ag}(m_a(A(\vec{x})),s,s') \supset AgtCanForce(m_a(A(\vec{x})),s)))
  \end{array}\]
\end{small}%
\end{constraint}

\noindent

Note that with only Constraint \ref{cstr:ProperMapp}, possibly combined with 
Constraint \ref{cstr:AgtAlwsCanExecuteHLactions}, there is no guarantee that when
a HL system action $A(\vec{x},e)$ is possibly executable, the environment can
actually ensure that reaction $e$ occurs when the agent executes $m(A(\vec{x}))$.
We can define an additional constraint to ensures this as follows.

First we define:

\begin{small}
\[\begin{array}{l} 
EnvCanForceBy(\delta, s, g) \doteq \forall P.[\ldots  \limp P(\delta, s)] \\
\mbox{ where \ldots stands for } \\
\lbrack (Final(\delta,s)) \land \lnot \exists s'.\exists \delta'. Trans(\delta,s,\delta',s')
\limp P(\delta, s)  \rbrack \land \\
\lbrack \exists A. \exists \vec{t}.\exists e.\exists \delta'. Trans(\delta,s,\delta',do(A(\vec{t},e),s))  \land \\
\forall A.\forall \vec{t}(\exists e.\exists \delta'.(Trans(\delta,s,\delta',do(A(\vec{t},e),s))) \limp \\
\exists e. \exists \delta'. g(A(\vec{t}),s) = e \land
    Trans(\delta,s,\delta',do(A(\vec{t},e),s))\\
    \qquad {} \land P(\delta',do(A(\vec{t},e),s))) \\
\limp P(\delta,s) \rbrack 
\end{array} \]
\end{small}%

\noindent
where $g$ is an environment strategy that maps situations and agent
actions to environment reactions.
Then we can require:

\begin{constraint} \label{EnvAlwsCanExecuteHLreactions}
(EnvAlwsCanExecuteHLreactions)\\
for every high-level system action
sequence $\vec{\alpha}$ and every high-level action $A$, we have that:

\begin{small}  
\[\begin{array}{l} 
  \D_l \cup \C \models \forall s. (Do(m_s(\vec{\alpha}),S_0,s) \limp\\
    \forall \vec{x},e.(\exists s'.Do(m_s(A(\vec{x},e)),s,s') \\
    \qquad \quad \supset \exists
    g. EnvCanForceBy(m_s(A(\vec{x},e)),s,g)))
  \end{array}\]
\end{small}%
\end{constraint}

\paragraph{Planning for Achievement Goals.}
Returning to planning, 
we can now show that if Constraint
\ref{cstr:AgtAlwsCanExecuteHLactions} holds
(i.e., the agent always knows how to execute high-level primitive
actions at the low level, even if not controlling their outcome)
and the agent has
a strong plan to achieve a goal at the high level, then there exists a
refinement of the high-level plan that is a strong plan to achieve the refinement of the
goal at the low level:

\begin{theorem} \label{thm:strongPlan} 
  If  $M_h \sim_m M_l$ and
  Constraint \ref{cstr:AgtAlwsCanExecuteHLactions} holds,
  then
  for any high-level system action sequence $\vec{\alpha}$ and any
  high-level situation-suppressed formula $\phi$,   we have that:

\begin{small}  
  \[\begin{array}{l}
            \mbox{if }  M_h,v \models Executable(do(\vec{\alpha},S_0))
      \land {}\\
      \hspace{5em} AgtCanForce(\phi, do(\vec{\alpha},S_0)) \\
      \mbox{then } M_l,v \models \exists s. Do(m_s(\vec{\alpha}),S_0,s)
      \land {} \\
      \hspace{5.9em} \forall s. Do(m_s(\vec{\alpha}),S_0,s) \limp  AgtCanForce(m_f(\phi),s)
\end{array}\]
\end{small}%
\end{theorem}

\noindent
\paragraph{Example Cont.}
As strategies use the $stop$ action, we first define a $stop_{HL}$
action at the high level which is mapped to the $stop_{LL}$ action at
the concrete level; $stop_{HL}$ always terminates with the $Success_{HS}$ reaction.
Observe that Constraint \ref{cstr:HLactionsNecTerminate} (and thus also the weaker
Constraint \ref{cstr:AgtAlwsCanExecuteHLactions}) is satisfied for our example domain (see appendix for details).

We can show that the agent has a strong plan to achieve the goal of
being at location $13$ at the high level ($\mathit{dtf}$ abbreviates $driveAndTryFix$ ):

\noindent
\begin{small}
\[\begin{array}{l}
\hspace{-0.5em}
\D_h^{tt} \models AgtCanForceBy(At_{HL}(13),S_0,	\mathit{f_h}) \mbox{    where } \\
\hspace{-0.5em}
	\mathit{f_h}(s) \doteq
    \begin{cases}
      stop_{HL} & \hspace{-0.5em} \text{if  } At_{HL}(13,s)\\
     	\mathit{dtf}(o,d) & \hspace{-0.5em} \text{if  }  At_{HL}(o,s) \land o \neq 13 \land Road_{HL}(o,d,s)   \\
			& \hspace{-0.5em} \land \neg 	\mathit{Visited_{HL}}(d,s) \land Spare_{HL}(d,s) \\
			wait_{HL} & \hspace{-0.5em} \text{otherwise}
    \end{cases}       
\end{array} \]
\end{small}%

\noindent
The agent's high-level strategy $\mathit{f_h}$ is to stop if she is already at location
$13$, otherwise to drive to a location which has not been visited
previously and has a spare tire and fix a flat if one occurs; in all other cases she waits.

By Th.~\ref{thm:strongPlan} there exists a strategy at the low level that is refinement of the high-level strategy; this strategy is the same as strategy $\mathit{f_l}$ in Section \ref{sec:backgroundCom}. It is easy to show that $\mathit{f_l}$ is a refinement of the high-level strategy $\mathit{f_h}$.

\paragraph{Planning for Temporally Extended Goals/Tasks.}
We can also show a similar result to the previous theorem for high-level
programs/tasks/temporally extended goals:
if Constraint \ref{cstr:AgtAlwsCanExecuteHLactions} holds and 
the agent has a strategy to successfully execute
an agent program (without concurrency) at the high level,
then the agent also has a strategy to successfully execute some
refinement of it at the low level:
\begin{theorem} \label{thm:agtCanForceProgram} 
  If  $M_h \sim_m M_l$ and Constraint \ref{cstr:AgtAlwsCanExecuteHLactions} holds,
 then for 
any high-level system action sequence $\vec{\alpha}$ and
 any SD \ConGolog\ high-level agent program $\delta$ without
 the concurrent composition construct,
  we have that:

\begin{small}  
  \[\begin{array}{l}
      \mbox{if }  M_h,v \models Executable(do(\vec{\alpha},S_0)) \land
      {}\\
      \hspace{5em} AgtCanForce(\delta, do(\vec{\alpha},S_0)) \\
      \mbox{then } M_l,v \models \exists s. Do(m_s(\vec{\alpha}),S_0,s)
      \land {} \\
      \hspace{5.9em} \forall s. Do(m_s(\vec{\alpha}),S_0,s) \limp  AgtCanForce(m_a(\delta),s)
\end{array}\]
\end{small}%
\end{theorem}

\noindent
Here, we apply the mapping to a high-level agent program $\delta$
without concurrency
to produce a low level agent program $m_a(\delta)$; this can be defined
in the obvious way, using $m_a$ to map atomic actions and $m_f$ for tests
as usual, and by mapping the components and composing the result for
other constructs, e.g.,
$m_a(\delta_1 \ndet \delta_2) = m_a(\delta_1) \ndet m_a(\delta_2)$

\noindent
\paragraph{Example Cont.}
Consider the high-level program
\begin{small} $\delta_{go}^{HL}(l) = (\pi o,d. \mathit{driveAndTryFix}(o,d))^*;
  At_{HL}(l)?$ \end{small}%
which goes to location $l$ by repeatedly picking adjacent locations
and doing $\mathit{driveAndTryFix}(o,d)$ until the agent is at $l$.
%
%
We can show that the agent has a strategy to successfully execute this
program to get to location $13$:

\noindent
\begin{small}
\[\begin{array}{l} 
\D^{tt}_h \models AgtCanForceBy(\delta_{go}^{HL}(13),S_0,f_h)
\end{array} \]
\end{small}%

\noindent
Here, the high-level strategy $\mathit{f_h}$ to execute the program is same as the
one we saw above.
It can also be shown that the low-level strategy $\mathit{f_l}$
seen earlier is a refinement of $f_h$ and can be used to to execute
$m_a(\delta_{go}^{HL}(13))$.
To further illustrate the use of high-level programs, notice that
we could give the agent the program
$\delta_{go}^{HL}(31); \delta_{go}^{HL}(13)$ to execute.  This is a
more complex task as she must first go to location $31$ and then to
$13$;
but it is easier to to find a strategy to do it.



%



\section{Example Involving Exogenous Events}

\begin{figure}[htp]
	\centering
		\includegraphics[scale=0.6]{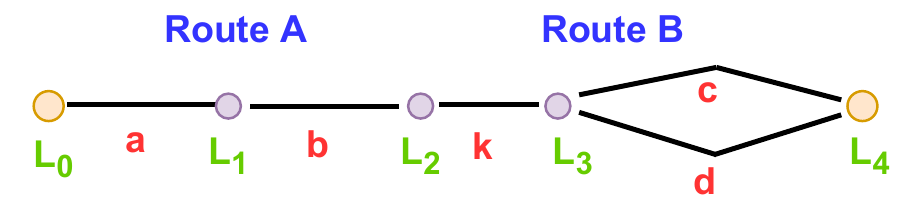}
	\caption{Logistics Domain Example}
	\label{fig:traverseRoutes}
	\vspace*{-2ex}
\end{figure}

Our framework can handle more complex domains including cases where we
have exogenous events.
These must be represented through environment reactions.
We illustrate this with a 
a simple logistics example where an agent, initially at location
$L_0$, needs to go to location $L_4$ (see Figure
\ref{fig:traverseRoutes}),
and where the exogenous events are roads/routes that open and close  (e.g., due to repairs).

\paragraph{High-Level NDBAT $\D_h^{r}$.}
At the high level, we abstract over navigation details. We assume we
have two routes, route $A$ $(Rt_A)$ going from $L_0$ to $L_2$ and
route $B$ $(Rt_B)$  going from $L_2$ to $L_4$. These
routes may get closed and may open again. For
simplicity, we assume that route status does not change while a route is
being taken (i.e., action $takeRoute$ is being performed);
route status may only change when the action
$checkRouteStatus$ is performed.\footnote{One can get a more realistic
  model by including the route status updates in the $takeRoute$
  environment reactions, and accommodating more outcomes, e.g.,
  having the agent return to the origin if a route is closes while she is trying to take it.}

We define a set of status updates for each route as follows: 


\begin{small}
\[\begin{array}{l} 
SetA_H = \{ close_{Rt}(Rt_A), open_{Rt}(Rt_A) \} 
\\
SetB_H = \{ close_{Rt}(Rt_B), open_{Rt}(Rt_B) \} 
\\
SetAB_H = SetA_H \cup SetB_H 
\end{array} \]
\end{small}%

\paragraph{Precondition Axioms}
$\D_h^{r}$ includes the following precondition axioms:

\noindent
\begin{small}
\[\begin{array}{l} 
Poss_{ag}(takeRoute(r, o, d),s) \doteq  o \neq d \land At_{HL}(o, s) \land \\
    \hspace*{1.5em} {}  CnRoute_{HL}(r, o, d,s) \land \neg Closed_{Rt}(r)
\\
 Poss_{ag}(checkRouteStatus,s) \doteq TRUE   
\\
\\
Poss(takeRoute(r, o, d, e_h), s) \equiv  
\\
\hspace*{1.5em} Poss_{ag}(takeRoute(r, o, d),s) \land 
\\
\hspace*{1.5em} \lbrack
(e_h = Success_{HL} \lor e_h = Failure_{HL})
    \rbrack
    \\
Poss(checkRouteStatus(e_h), s) \equiv  
\\
\hspace*{1.5em} Poss_{ag}(checkRouteStatus,s) \land 
\\
\hspace*{1.5em} \lbrack
((e_h \subset SetAB_H) 
\land
\\
\hspace*{1.5em} \forall r. \neg (close_{Rt}(r) \in e_h \land open_{Rt}(r) \in e_h)
\rbrack    
\end{array} \]
\end{small}%

The system action $takeRoute(r, o, d, e_h)$ can be performed by the agent to go from origin location $o$ to destination
location $d$ via route $r$ (see Figure \ref{fig:traverseRoutes}), and $e_h$ models the environment reaction. This action is executable when the agent action $takeRoute(r, o, d)$ is executable (i.e., the agent is initially at $o$, route $r$ connects $o$ to $d$, $r$ is not closed). $e_h$ could either represent success or failure of taking the route.

The system action $checkRouteStatus(e_h)$ can be performed to check
the status of routes and $e_h$ represents the environment
reaction. This action is executable when the agent action
$checkRouteStatus$ is executable (which is always) and $e_h$ could be
any subset of route opening and closures as long as opening a certain
route and closing the same route do not occur
simultaneously. \footnote{Here
  and below, we use set notation for simplicity.
Since we can't use function symbols, we need to encode membership of
status updates in reactions using predcates such as $ContainsOpen_{Rt}(e_h,r,s)$.}

\paragraph{SSAs}
$\D_h^{r}$ includes the following successor state axioms:

\noindent
\begin{small}
\[\begin{array}{l} 
At_{HL}(l, do(a,s)) \equiv \\
\hspace*{1.5em} (\exists l', e_h, r. a = takeRoute(r, l', l, e_h) \land 
(e_h=Success_{HL}))   
\\
\hspace*{1.5em} \lor 
\\
\hspace*{1.5em}(At_{HL}(l, s) \; \land \\
\hspace*{1.5em} \forall l', r, e_h. a \neq  takeRoute(r, l, l', e_h) \land e_h=Success_{HL})
\\
\\
Closed_{Rt}(r, do(a,s)) \equiv \\
\hspace*{1.5em} \exists e_h. a = checkRouteStatus(e_h)
\land 
close_{Rt}(r) \in e_h \\
\hspace*{1.5em} \lor 
\\
\hspace*{1.5em} \lbrack Closed_{Rt}(r, s) \land \\
\hspace*{1.5em} \forall e'_h. a \neq checkRouteStatus(e'_h) \land open_{Rt}(r) \in e'_h 
\rbrack
\end{array} \]
\end{small}%

\noindent
For the other fluents, we have SSAs specifying that they are unaffected by any actions.

\paragraph{Initial State Axioms}
$\D_h^{r}$ also contains the following initial state axioms:  

\begin{small}
\[\begin{array}{l} 
At_{HL}(l, S_0) \equiv l=L0, \\
\forall r. \neg Closed_{Rt}(r,S_0), \\
CnRoute_{HL}(r, o, d, S_0) \equiv \\
\hspace*{1.5em} (r,o,d) \in \{(Rt_A, L0, L2), (Rt_A, L2, L0),\\
\hspace*{6.5em} (Rt_B, L2, L4), (Rt_B, L4, L2)\}
\end{array} \]
\end{small}%

\noindent
So initially, it is assumed that all of the routes are open.

\paragraph{Low-Level NDBAT $\D_l^{r}$.}
At the low level, we model navigation in a more detailed way. The agent has a more detailed map with more locations and roads between
them. These roads may get closed and may open again. Again, for
simplicity, we assume road status does not change while a road is
being taken (i.e., action $takeRoad$ is being performed); road status may only change when the action $checkRoadStatus$ is performed.

We keep a separate set of status updates for roads that constitute a route in the high level as follows: 


\begin{small}
\[\begin{array}{l} 
SetA_L = \{ close_{Rd}(Rd_a), close_{Rd}(Rd_b), 
\\
\quad open_{Rd}(Rd_a), open_{Rd}(Rd_b) \} 
\\
SetB_L = \{ close_{Rd}(Rd_k), close_{Rd}(Rd_c), 
close_{Rd}(Rd_d), \\
\quad open_{Rd}(Rd_k), open_{Rd}(Rd_c), open_{Rd}(Rd_d) \} 
\\
SetAB_L = SetA_L \cup SetB_L 
\end{array} \]
\end{small}%

\paragraph{Precondition Axioms}
$\D_h^{r}$ includes the following precondition axioms:

\noindent
\begin{small}
\[\begin{array}{l} 
Poss_{ag}(takeRoad(t, o, d),s) \doteq  o \neq d \land At_{LL}(o, s) \land \\
\hspace*{7em} {}  CnRoad(t, o, d,s) \land \neg Closed_{Rd}(t)
    \\
Poss_{ag}(checkRoadStatus,s) \doteq TRUE
\\[1ex]
Poss(takeRoad(t, o, d, e_l), s) \equiv  
\\
\hspace*{1.5em} Poss_{ag}(takeRoad(t, o, d),s) \land 
\\
\hspace*{1.5em} \lbrack
(e_l = Success_{LL} \lor e_l = Failure_{LL})
    \rbrack
    \\
    Poss(checkRoadStatus(e_l), s) \equiv  
\\
\hspace*{1.5em} Poss_{ag}(checkRoadStatus,s) \land 
\\
\hspace*{1.5em} \lbrack
((e_l \subset SetAB_L) 
\land
\\
\hspace*{1.5em} \forall t. \neg (close_{Rd}(t) \in e_l \land open_{Rd}(t) \in e_l)
\rbrack
\end{array} \]
\end{small}%

\noindent
Thus the system actions $checkRoadStatus(e_l)$ and $takeRoad(t,o,d,e_l)$ are defined similarly to $checkRouteStatus(e_h)$ and $takeRoute(r,o,d,e_h)$ above, with the difference that these actions consider roads instead of routes and environment reactions model closure and opening of roads.

\paragraph{SSAs}
$\D_l^{r}$ includes the following successor state axioms:

\noindent
\begin{small}
\[\begin{array}{l} 
At_{LL}(l, do(a,s)) \equiv \\
\hspace*{1.5em} (\exists l', e_l, t. a = takeRoad(t, l', l, e_l) \land 
(e_l=Success_{LL}))   
\\
\hspace*{1.5em} \lor 
\\
\hspace*{1.5em} (At_{LL}(l, s) \; \land \\
\hspace*{1.5em} \forall l', t, e_l. \\
\hspace*{1.5em} (a \neq  takeRoad(t, l, l', e_l) \land e_l=Success_{LL}))
\\
\\ 
Closed_{Rd}(t, do(a,s)) \equiv \\
\hspace*{1.5em} \exists e_l. a = checkRoadStatus(e_l)
\land 
close_{Rd}(t) \in e_l
\\
\hspace*{1.5em} \lor \\
\hspace*{1.5em} \lbrack Closed_{Rd}(t, s) \land \\
\hspace*{1.5em} \forall e'_l. 
a \neq checkRoadStatus(e'_l) \land 
open_{Rd}(t) \in e'_l 
\rbrack
\end{array} \]
\end{small}%

\noindent
For the other fluents, we have SSAs specifying that they are unaffected by any actions.

\paragraph{Initial State Axioms}
$\D_l^{r}$ also contains the following initial state axioms:  

\begin{small}
\[\begin{array}{l} 
At_{LL}(l, S_0) \equiv l=L0, \\
\forall t. \neg Closed_{Rd}(t,S_0),\\ 
CnRoad(t, o, d, S_0) \equiv \\
\hspace*{1.5em} (t,o,d) \in \{(Rd_a, L0, L1), (Rd_a, L1, L0), \\
\hspace*{6.5em} (Rd_b, L1, L2),(Rd_b, L2, L1), \\
\hspace*{6.5em} (Rd_k, L2, L3),(Rd_k, L3, L2), \\
\hspace*{6.5em} (Rd_c, L3, L4), (Rd_c, L4, L3), \\
\hspace*{6.5em} (Rd_d, L3, L4), (Rd_d, L4, L3) \}
\end{array} \]
\end{small}%

\noindent
together with a complete specification of $CnRoute_{LL}$. Initially, it is assumed that none of the roads are closed.

\paragraph{Saving the State of Road Closures}
The route status updates that should occur when the high-level
action $checkRouteStatus$ is done depend on the changes in the 
status of the associated roads that occur when the low-level action
$checkRoadStatus$ is done.
To define the mapping for $checkRouteStatus$,  it is useful to introduce a new low-level action $saveClosedRdState(e_l)$ that memorizes the
values of fluent $Closed_{Rd}(t)$ at the low level. First,
we add a new fluent $Closed^{Start}_{Rd}(t)$ (where $t$ is a road) to save the status of the low-level fluent $Closed_{Rd}(t)$ prior to execution of a refinement of a high-level action. Then the extension of $Closed_{Rd}(t)$ is copied into $Closed^{Start}_{Rd}(t)$ when action $saveClosedRdState$ is performed.  We can then compare status of closure of roads before and after a refinement of $checkRouteStatus$ has been executed.

We formalize this by adding the following axioms to the low-level theory:

\begin{small}
\[\begin{array}{l}
 Closed^{Start}_{Rd}(t,do(a,s)) \equiv \\
   \hspace*{1.5em}  (\exists e_l.a = saveClosedRdState(e_l) \land Closed_{Rd}(t,s)
    \\ 
		\hspace*{1.5em} \lor \\ 
		\hspace*{1.5em} Closed^{Start}_{Rd}(t,do(a,s)) \land \forall e_l. a \neq saveClosedRdState(e_l))
		\\[1ex]
		Poss_{ag}(saveClosedRdState,s) \doteq True
		\\
		Poss(saveClosedRdState(e_l),s) \equiv \\
		\hspace*{1.5em} Poss_{ag}(saveClosedRdState,s) \land e_l=Success_{CL}

    \\[1ex]
    \forall t. \lnot  Closed^{Start}_{Rd}(t,S_0)
\end{array}\]
\end{small}%


\noindent
We also define the following abbreviations: 

\begin{small}
\[\begin{array}{l}
Closed^{Start}_{Rt}(Rt_A) \doteq  Closed^{Start}_{Rd}(Rd_a) \lor Closed^{Start}_{Rd}(Rd_b)\\
Closed^{Start}_{Rt}(Rt_B) \doteq  Closed^{Start}_{Rd}(Rd_k) \lor \\
\qquad {} (Closed^{Start}_{Rd}(Rd_c) \land Closed^{Start}_{Rd}(Rd_d)) \\
\end{array}\]
\end{small}%

\paragraph{Refinement Mapping $m^{r}$.}
We specify the relationship between the high-level and low-level
NDBATs through the following refinement mapping $m^r$.

The fluents are mapped as follows. 

\begin{small}
\[\begin{array}{l} 
m^{r}_f(At_{HL}(l))= At_{LL}(l) 
\\
\\
m^{r}_f(Closed_{Rt}(r))=\\
\hspace*{1em} r=Rt_A \land (Closed_{Rd}(Rd_a) \lor Closed_{Rd}(Rd_b)) \; \lor
\\
\hspace*{1em} r=Rt_B \land (Closed_{Rd}(Rd_k) \lor \\
\hspace*{1em} \qquad {} (Closed_{Rd}(Rd_c) \land Closed_{Rd}(Rd_d)))
\\
m^{r}_f(CnRoute_{HL}(r,o,d)) = CnRoute_{LL}(r,o,d)
\end{array} \]
\end{small}%

\noindent
Thus for example, route $Rt_B$ is closed if either road $Rd_k$ is closed or both roads $Rd_c$ and $Rd_d$ are closed. 

For the agent action $takeRoute(r, o, d)$ we have:

\begin{small}
\[\begin{array}{l} 
m^{r}_a(takeRoute(r, o, d))= \\
\hspace*{1em} CnRoute_{LL}(r,o,d)?;\\
\hspace*{1em} (\exists t_1,l,t_2. CnRoad(t_1,o,l) \land \neg Closed_{Rd}(t_1) \land \\
\hspace*{1em} CnRoad(t_2,l,d) \land \neg Closed_{Rd}(t_2))?; \\
\hspace*{1em} \lbrack
\pi t_1,t_2, l.(takeRoad(t_1, o, l); At_{LL}(l)?;\\
\hspace*{2em} takeRoad(t_2, l, d); At_{LL}(d)?)
\rbrack  \;\; \mid{} 
\\
\hspace*{1em} \lbrack
\pi t_1,t_2, l. ((takeRoad(t_1, o, l); At_{LL}(o)?)  \mid \\
\hspace*{2em} (takeRoad(t_1, o, l);  At_{LL}(l)?;\\
\hspace*{2.5em}  takeRoad(t_2, l, d); At_{LL}(l)?; \\
    \hspace*{2.5em} takeRoad(t_1, l, o); At_{LL}(o)?))
    \rbrack
\end{array} \]
\end{small}%


\noindent
Note that we assume that if the second leg of a route cannot be
taken, then the agent will be able to return to the origin.

For the system action $takeRoute(r, o, d, e_h)$ we have:

\begin{small}
\[\begin{array}{l} 
m^{r}_s(takeRoute(r, o, d, e_h))= \\
\hspace*{1em} CnRoute_{LL}(r,o,d)?;\\
\hspace*{1em} (\exists t_1,l,t_2. CnRoad(t_1,o,l) \land \neg Closed_{Rd}(t_1) \land {}\\
\hspace*{2em} CnRoad(t_2,l,d) \land \neg Closed_{Rd}(t_2))?; \\
    \hspace*{1em} \lbrack
\pi t_1,t_2, l.(\pi e_l.takeRoad(t_1, o, l,e_l); At_{LL}(l)?;\\
\hspace*{2em} \pi e_l.takeRoad(t_2, l, d,e_l); At_{LL}(d)?);
\\
\hspace*{2em} e_h=Success_{HL}? \rbrack  \;\; \mid{} 
    \\
    \hspace*{1em} \lbrack
\pi t_1,t_2, l. (\pi e_l (takeRoad(t_1, o, l, e_l); At_{LL}(o)?)  \mid \\
    \hspace*{2em} (\pi e_l. takeRoad(t_1, o, l,e_l);  At_{LL}(l)?;\\
    \hspace*{2.5em} \pi e_l. takeRoad(t_2, l, d,e_l); At_{LL}(l)?;\\
\hspace*{2.5em} \pi e_l. takeRoad(t_1, l, o,e_l); At_{LL}(o)?)); 
\\
    \hspace*{2em} e_h=Failure_{HL}? \rbrack
\end{array} \]
\end{small}%


For the agent action $checkRouteStatus$ we have:

\begin{small}
\[\begin{array}{l} 
m^{r}_a(checkRouteStatus)= \\
\hspace*{1em} \pi e_l. saveClosedRdState(e_l);
    \pi e_l. checkRoadStatus(e_l);
\end{array} \]
\end{small}%

For the system action $checkRouteStatus(e_h)$ we have:

\begin{small}
\[\begin{array}{l} 
m^{r}_s(checkRouteStatus(e_h))= \\
\hspace*{1em} \pi e_l. saveClosedRdState(e_l);
    \pi e_l. checkRoadStatus(e_l);
\\
\hspace*{1em} (\forall r. (\exists o,d.CnRoute_{LL}(r,o,d)) \limp
\\
\hspace*{1.3em} \lbrack close_{Rt}(r) \in e_h \equiv
\neg Closed^{start}_{Rt}(r) \land m_f(Closed_{Rt}(r))
\\
\hspace*{1.3em} \land 
\\
\hspace*{1.3em} open_{Rt}(r) \in e_h \equiv 
Closed^{start}_{Rt}(r) \land  m_f(\neg Closed_{Rt}(r))
\rbrack)? 
\end{array} \]
\end{small}%

\noindent
Note how we include in the high-level reaction $e_h$ status updates
for all the routes whose status has changed.





%
\section{Monitoring and Explanation}
In many application contexts, one is interested in tracking/monitoring what
the low-level agent is doing and describing it in abstract terms
(e.g., to a client or manager). This is particularly important in the
context of \emph{Explainable AI}.
In this section, we review the monitoring and explanation definitions
and results for BAT abstractions of (BDL17) and show that they can be
carried forward to NDBAT abstractions.  This follows essentially from
the fact that we have not changed the definition of $m$-bisimulation
from (BDL17), but are applying it here to sequences of system actions.

First, let's recall the following definition from (BDL17): $\D_h$ is a \emph{sound abstraction of} $\D_l$ \emph{relative to refinement mapping} $m$ if and only if, for all models $M_l$ of $\D_l \cup \C$, there exists a model $M_h$ of $\D_h$ such that  $M_h \sim_m M_l$.
Suppose we have a ground low-level situation
term $S_l$ such that $\D_l \cup \{ Executable(S_l) \}$ is satisfiable, 
and $\D_l \cup \{ Do(m_s(\vec{\alpha}),S_0,S_l) \}$ is satisfiable, then
the ground high-level system action sequence $\vec{\alpha}$ is a possible way of
describing in abstract terms what has occurred in getting to situation
$S_l$.  

If $\D_h \cup \{ Executable(do(\vec{\alpha},S_0)) \}$ is also
satisfiable (it must be if $\D_h$ is a sound abstraction of $\D_l$ wrt
$m$), then one can also answer high-level queries about what may hold
in the resulting situation, i.e., whether
$\D_h \cup \{ Executable(do(\vec{\alpha},S_0)) \land
\phi(do(\vec{\alpha},S_0))\}$
is satisfiable, and what must hold in such a resulting situation, i.e.,
whether $\D_h \cup \{ Executable(do(\vec{\alpha},S_0))\} \models
\phi(do(\vec{\alpha},S_0))$.  
One can also answer queries about
what high-level action $\beta$ might occur next,
 i.e., whether $\D_h \cup \{ Executable(do(\vec{\alpha}\beta,S_0))\}$
is satisfiable.

Note that in general, there may be several distinct ground high-level
system action sequences $\vec{\alpha}$ that match a ground low-level
situation term $S_l$; even if we have complete information and a
single model $M_l$ of $\D_l \cup \C$, there may be several distinct
$\vec{\alpha}$'s such that
$M_l \models Do(m_s(\vec{\alpha}),S_0,S_l)$.

\paragraph{Example.} Suppose that we have two high-level system actions $a_1(r_{A1})$ and $a_2(r_{A2})$ where the environment reactions $r_{A1}$ and $r_{A2}$ take on the values $R_{A1}$ and $R_{A2}$ respectively. The refinement mapping is defined as follows: $m_s(a_1(r_{A1}))= (c(R_{C}) \mid d(R_{D}));r_{A1}=R_{A1}?$ and $m_s(a_2(r_{A2}))= (d(R_{D}) \mid e(R_{E}));r_{A2}=R_{A2}?$ where $R_{C}$, $R_{D}$, and $R_{E}$ represent the environment reactions to low level actions $c$, $d$, and $e$ respectively. Assume all actions are always executable. We have 
that $M_h \models Executable(do(a_1(R_{A1})),S_0)$, $M_h \models Executable(do(a_2(R_{A2})),S_0)$, and $M_l \models Executable(do(d(R_{D})),S_0)$. 
Then the low-level situation $do(d(R_{D}), S_0)$ can be reached by
performing both $Do(m_s(a_1(R_{A1})),S_0)$ and
$Do(m_s(a_2(R_{A2}),S_0)$. 

In many contexts, this would be counterintuitive and we would like to
be able to map a sequence of low-level actions performed by the
low-level agent back into a \emph{unique} abstract high-level action
sequence it refines, i.e., we would like to define an inverse mapping
\emph{function} $m_s^{-1}$.  To do so, we adapt definitions, assumptions and results originally introduced by (BDL17) on monitoring to (sequences of) system actions in NDBATs. In the following, we assume that the refinement mapping $m$ is proper wrt $\D_l$.

We first define some low-level system programs that characterize
the refinements of high-level system action/action sequences: 

\begin{small}
\[\begin{array}{l}
\mathit{\anyonehlon} \doteq \ \ \mid_{A_i \in \A_h} \pi \vec{x} \pi e.m_s(A_i(\vec{x},e))
    \\ 
    \qquad ~\mbox{i.e., do any refinement of any HL primitive system action,}
\\[1ex]
\mathit{\anyseqhl} \doteq \mathit{\anyonehlon}^*
\\ 
\qquad ~~\mbox{i.e., do any sequence of refinements of HL system actions.}
\end{array}\]
\end{small}%

We start by adapting the abbreviation $lp_{m}(s,s')$ from (BDL17). The abbreviation $lp_{m}(s,s')$ 
means that $s'$ is a largest prefix of $s$
that can be produced by executing a refinement of a sequence of high-level system actions:

\begin{small}
\[\begin{array}{l}
\hspace*{2.5em}lp_m(s,s') \doteq
Do(\anyseqhl,S_0,s') \land {}
s' \leq s \land {}\\
\hspace*{4.5em} \forall s''. (s' < s'' \leq s \supset \lnot Do(\anyseqhl,S_0,s''))
\end{array}\]
\end{small}%

\noindent
(BDL17) show that the relation $lp_m(s,s')$ is actually a total function. 
Given this result, we can introduce the notation $lp_m(s) = s'$ to
stand for $lp_m(s,s')$.





To be able to map a low-level system action sequence back to a unique
high-level system action sequence that produced it, we need the following
assumption:

\begin{assumption} \label{asm:firstThree}
For any distinct ground high-level system action terms $\alpha$ and $\alpha'$ we have that:
 
\begin{small}
\begin{description}
\item[(a)] $\D_l \cup \C \models \forall s,s'. Do(m_s(\alpha), s,s') \limp$ \\
\hspace*{9em} $ \neg \exists \delta . Trans^*(m_s(\alpha'),s,\delta,s')$

\item[(b)] 
$\D_l \cup \C \models \forall s,s'. Do(m_s(\alpha), s,s') \limp$ \\
\hspace*{9em} $ \neg \exists a \exists \delta. Trans^*(m_s(\alpha),s,\delta,do(a,s'))$

\item[(c)] 
$\D_l \cup \C \models  \forall s,s'. Do(m_s(\alpha), s,s') \limp  s < s'$

\end{description}
\end{small}
\end{assumption}
\noindent
Part $(a)$ ensures that different high-level primitive system actions
have disjoint sets of refinements; $(b)$ ensures that once a refinement of a 
high-level primitive system action is complete, it cannot be extended further; and $(c)$ ensures
that a refinement of a high-level primitive system action will produce at least one low-level action.

The three conditions in Assumption \ref{asm:firstThree} ensure that if we have a low-level system action
sequence that can be produced by executing some high-level system action
sequence, there is a unique high-level action sequence that can produce it:

\begin{theorem}[BDL17]\label{thm:unique-m_inv}
  Suppose that we have a refinement mapping $m$ from $\D_h$ to $\D_l$
  and that Assumption \ref{asm:firstThree} holds. Let
  $M_l$ be a model of $\D_l \cup \C$. Then for any ground situation
  terms $S_s$ and $S_e$ such that
  $M_l \models Do(\mathit{\anyseqhl},S_s, S_e)$, there exists a unique
  ground high-level system action sequence $\vec{\alpha}$ such that
  $M_l \models Do(m_s(\vec{\alpha}),S_s,S_e)$.
\end{theorem}

Since in any model $M_l$ of $\D_l \cup \C$, for any ground situation
term $S$, there is a unique largest prefix of $S$ that can be produced
by executing a sequence of high-level system actions, $S'=lp_m(S)$, and for
any such $S'$, there is a unique ground high-level system action
sequence $\vec{\alpha}$ that can produce it, we can view
$\vec{\alpha}$ as the value of the inverse mapping $m_s^{-1}$ for $S$ in
$M_l$.

\begin{small}
\[\begin{array}{l}
m^{-1}_{M_l}(S) = \vec{\alpha} \doteq 
M_l \models \exists s'. lp_m(S)= s' \land
    Do(m_s(\vec{\alpha}),S_0,s')
\mbox{}
  \end{array}\]
\end{small}%
\noindent  
  where $m_s$ is a refinement mapping from $\D_h$ to $\D_l$ and
  Assumption \ref{asm:firstThree} holds, $M_l$ is a model of
  $\D_l \cup \C$, $S$ is a ground low-level situation term, and
  $\vec{\alpha}$ is a ground high-level system action sequence.

Assumption \ref{asm:firstThree} however does not 
ensure that any low-level situation $S$ can 
in fact be generated by executing a refinement of some
high-level system action sequence; if it cannot, then the inverse mapping
will not return a complete matching high-level system action sequence (e.g., we might have
$m^{-1}_{M_l}(S) = \epsilon$).

We can introduce an additional assumption that rules this
out:\footnote{One might prefer a weaker version of
Assumption \ref{asm:offMatchLL}.
For instance,  one could write a program
specifying the low level agent's possible behaviors and require that
situations reachable by executing this program can be generated by
executing a refinement of some high-level system action sequence.}

\begin{assumption} \label{asm:offMatchLL}
~~\\[0.5ex]
\begin{small}
$D_l \cup \C \models \forall s. \mathit{Executable}(s) \limp
\exists \delta.Trans^*(\anyseqhl,S_0,\delta,s)$
\end{small}%
\end{assumption}

With this additional assumption, we can show 
that for any executable low-level situation $s$, what remains after
the largest prefix that can be produced by executing a sequence of
high-level actions, i.e., the system actions in the interval between $s'$ and
$s$ where $lp_m(s,s')$, can 
be generated by some (not yet
complete) refinement of a high-level primitive system action:

\begin{theorem}[BDL17]\label{thm:lp_m_rest}
If  $m$ is a refinement mapping from $\D_h$ to $\D_l$ and 
Assumption \ref{asm:offMatchLL} holds, then we have that:

\begin{small}
\[\begin{array}{l}
\D_l \cup \C \models \forall s, s'. Executable(s) \land lp_m(s,s')
\supset \\
\hspace{6.6em}\exists \delta.Trans^*(\anyonehlon,s',\delta,s)
\end{array}\]
\end{small}%
\end{theorem}



\paragraph{Example.}
Let us consider an extended version of our triangle tire world example. In
this version, if there is no spare tire in the location where the car
is at (i.e., locations $11$, $12$ and $13$), the agent can always buy
a tire to replace the flat, or alternatively, if the location is
covered  by a roadside assistance service that the agent has
subscribed to, then she can use this service to fix the flat tire.


\paragraph{NDBAT $\D_l^{tt+}$.} We start by defining actions that are new in this variation of the example. The system action $order(l,r)$ lets the agent order a tire at location $l$ and is executable when the agent is at $l$, there is a flat tire, no spare tire is available at $l$ and a tire has not been already ordered. The environment reaction $r$ is $Succ_{LOrder}$. 
The fluent $Ordered(l,s)$ indicates that a tire has been ordered at location $l$. The system action $pay(l,r)$ lets the agent pay for an ordered tire at location $l$, and it can be performed when the agent has already placed an order but has not paid yet; $r$ is $Succ_{LPay}$. 
The fluent $Paid(l,s)$ indicates that the ordered item has already been paid for. The system action $callService(l,r)$ allows the agent to call for service at location $l$, and environment's reaction $r$ is $Succ_{LServe}$. This action can be performed when the agent is at location $l$, there is a flat tire,  no spare tire is available at $l$, and $l$ falls under the service coverage area and the agent has not already called for service.  The fluent $CalledServ(l,s)$ indicates that service has been requested at location $l$. 
 
Similar to before, the system action $fixFlatTire(l,r)$ fixes the flat
tire in location $l$, and can be performed when the agent is at
location $l$ and there is a flat tire, and either a spare tire is
available at $l$, or, the agent has already called for service or
ordered and paid for the tire. The environment reaction $r$ is
$Success_{LF}$. The system action $drive(o,d,r)$, and fluents
$At_{LL}(l,s)$, $Road_{LL}(o,d,s)$, $Spare_{LL}(l,s)$, and
$Flat_{LL}(s)$ are axiomatized  exactly as before (see Section \ref{sec:backgroundCom}).

$\D_l^{tt+}$ includes the following action precondition axioms:


\begin{small}
\[\begin{array}{l} 
Poss_{ag}(fixFlatTire(l), s) \doteq  \\
\hspace*{1.3em} At_{LL}(l,s) \land Flat_{LL}(s) \land  {[} Spare_{LL}(l,s) \lor \\
\hspace*{1.3em}  (\neg Spare_{LL}(l,s) \land (Paid(l,s) \lor CalledSrv(l,s))) {]} 
\\
Poss_{ag}(order(l), s) \doteq  \\
\hspace*{1.3em}  At_{LL}(l,s) \land \neg Spare_{LL}(l,s) \land Flat_{LL}(s) \land \neg Ordered(l,s)
\\
Poss_{ag}(pay(l), s) \doteq  \\
    \hspace*{1.3em} Ordered(l,s) \land \neg Paid(l,s)
    \\
Poss_{ag}(callService(l), s) \doteq  \\
\hspace*{1.3em} At_{LL}(l,s) \land \neg Spare_{LL}(l,s) \land Flat_{LL}(s) \land SrvZone_{LL}(l,s)  \\
\hspace*{1.3em} \land \neg CalledSrv(l,s)
\\[1ex]
Poss(fixFlatTire(l,r), s) \equiv  \\
\hspace*{1.3em} Poss_{ag}(fixFlatTire(l), s) \land r=Success_{LF} 
\\
Poss(order(l,r), s) \equiv  \\
\hspace*{1.3em} Poss_{ag}(order(l), s) \land r=Succ_{LOrder}
\\
Poss(pay(l,r), s) \equiv  \\
\hspace*{1.3em} Poss_{ag}(pay(l), s) \land r=Succ_{LPay} 
\\
Poss(callService(l,r), s) \equiv  \\
\hspace*{1.3em} Poss_{ag}(callService(l), s) \land r=Succ_{LServ} 
\end{array} \]
\end{small}%

$\D_l^{tt+}$ also includes the following SSAs:

\begin{small}
\[\begin{array}{l} 
Ordered(l,do(a,s)) \equiv  \exists r. a=order(l,r) \lor \\
\hspace*{1.3em}  Ordered(l,s) \land \forall r. a \neq fixFlatTire(l,r)
\\
Paid(l,do(a,s)) \equiv \exists r. a=pay(l,r) \lor \\
\hspace*{1.3em} Paid(l,s) \land \forall r.a \neq fixFlatTire(l,r)    
\\ 
CalledSrv(l,do(a,s)) \equiv \exists r. a=callService(l,r) \lor \\
\hspace*{1.3em} CalledSrv(l,s) \land \forall r. a \neq fixFlatTire(l,r) 
\end{array} \]
\end{small}%

\noindent 
For the $SrvZone_{LL}(l,s)$ fluent, we have a SSA specifying that it is unaffected by any action.


$\D_l^{tt+}$  also contains the following initial state axioms:

\noindent
\begin{small}
\[\begin{array}{l} 
Road_{LL}(o,d,S_0) \equiv (o,d) \in \{ (11,12), (11,21), (12,21),\\
(12,22), (12,13),(13,22),(22,31),(21,31), (12,11), (21,11),\\
(21,12), (22,12), (13,12), (22,13),(31,22),(31,21) \}, \\
Spare_{LL}(l,S_0) \equiv l \in \{21,22,31\}, At_{LL}(l,S_0) \equiv l=11, \\
SrvZone_{LL}(l,S_0) \equiv l \in \{12,13\}, \neg Flat_{LL}(S_0),\\
\forall l. \neg Paid(l,S_0), \forall l. \neg Ordered(l,S_0), \forall l. \neg CalledSrv(l,S_0).\\
\end{array} \]
\end{small}%

Note that we have complete information at the low level and a single model of $\D_l^{tt+}$. Also note that 
in the initial situation, it is known that only locations $12$ and $13$ are considered to be service zones.

\paragraph{NDBAT $\D_h^{tt+}$.} We define a high-level NDBAT $\D_h^{tt+}$
that abstracts over driving and fixing potential flat tires, purchasing and fixing flat tires, as well as requesting service to fix a flat tire.
The system action $driveAndTryFix(o,d,r)$, as well as fluents $At_{HL}(l,s)$, $Road_{HL}(o,d,s)$, and $Spare_{HL}(l,s)$  are axiomatized exactly as before (see Section \ref{sec:NDabstraction}). The system action $buyAndFix(l,r)$ can be performed to buy a tire and repair the flat tire, and is executable when there is a flat tire, and the agent is at location $l$ where a spare is not available. $r$ can take on the value $Succ_{HBuy}$.  
The system action $serviceAndFix(l,r)$ lets the agent call for service
and repair the flat tire and is executable when  there is a flat tire,
and the agent is at location $l$ where a spare is not available and
$l$ is a service zone. The environment reaction $r$ is $Succ_{HServ}$.

$\D_h^{tt+}$ includes the following system and agent action precondition axioms:

\noindent
\begin{small}
\[\begin{array}{l} 
Poss_{ag}(buyAndFix(l), s) \doteq  \\
\hspace*{1.3em} At_{HL}(l,s) \land \neg Spare_{HL}(l,s) \land Flat_{HL}(s) 
\\
Poss_{ag}(serviceAndFix(l), s) \doteq  \\
\hspace*{1.3em}  At_{HL}(l,s) \land \neg Spare_{HL}(l,s) \land Flat_{HL}(s) \land \\
\hspace*{1.3em}  SrvZone_{LL}(l,s) 
\\[1ex]
Poss(buyAndFix(l,r), s) \equiv \\
\hspace*{1.3em} Poss_{ag}(buyAndFix(l), s) \land r=Succ_{HBuy} 
\\
Poss(serviceAndFix_{HL}(l,r), s) \equiv  \\
\hspace*{1.3em} Poss_{ag}(serviceAndFix(l), s) \land r=Succ_{HServ} 
\end{array} \]
\end{small}%

$\D_h^{tt+}$ also includes the following SSA:

\noindent
\begin{small}
\[\begin{array}{l}
Flat_{HL}(do(a,s)) \equiv \\
\hspace*{1.3em} \exists o, d. a=driveAndTryFix(o,d, DrvFlat) \; \lor \\
\hspace*{1.3em} Flat_{HL}(s) \land \forall r_{h1},r_{h2}. a \neq buyAndFix(l,r_{h1}) \land \\
\hspace*{1.3em} a \neq serviceAndFix(l,r_{h2})
\end{array} \]
\end{small}%

\noindent
For the $SrvZone_{HL}(l,s)$ fluent, we have a SSA specifying that it is unaffected by any action.


$\D_h^{tt+}$  also contains the following initial state axioms:

\noindent
\begin{small}
\[\begin{array}{l}
Road_{HL}(o,d,S_0) \equiv (o,d) \in \{ (11,12), (11,21), (12,21),\\
(12,22), (12,13),(13,22),(22,31),(21,31), (12,11), (21,11),\\
(21,12), (22,12), (13,12), (22,13),(31,22),(31,21) \}, \\
Spare_{HL}(l,S_0) \equiv l \in \{21,22,31\}, At_{HL}(l,S_0) \equiv l=11,\\
\neg Flat_{HL}(S_0), SrvZone_{HL}(13,S_0).
\end{array} \]
\end{small}%

Note that in the initial situation, the agent only knows that location $13$ is considered a service zone, and the agent does not have know whether $11$ or $12$ are service zones as well. Hence, there is more than one model of $\D_h^{tt+}$. 

\paragraph{Refinement Mapping $m^{tt+}$}
We specify the relationship between the high-level and low-level NDBATs through the following refinement mapping $m^{tt+}$:

\begin{small}
\[\begin{array}{l}
m^{tt+}_a(buyAndFix(l)) = order(l);pay(l);fixFlatTire(l)
\\
m^{tt+}_a(serviceAndFix(l)) = callService(l);fixFlatTire(l)
\\
\\

m^{tt+}_s(buyAndFix(l,r_h))= \\
\hspace*{1.3em} order(l,Succ_{LOrder}); pay(l,Succ_{LPay});\\
\hspace*{1em} fixFlatTire(l,Success_{LF});r_h=Succ_{HBuy}?
\\
m^{tt+}_s(serviceAndFix(l,r_h))= \\
\hspace*{1.3em} callService_{LL}(l,Succ_{LServ});fixFlatTire(l,Success_{LF}); \\
\hspace*{1.3em} r_h=Succ_{HServ}?
\\
\\
m^{tt+}_f(SrvZone_{HL}(l))=SrvZone_{LL}(l)
\end{array} \]
\end{small}%

Thus, the high-level agent actions $serviceAndFix(l)$ and
$buyAndFix(l)$ are refined by low-level programs where the agent
performs the sequences of agent actions $callService(l);fixFlatTire(l)$ and $order(l);pay(l);fixFlatTire(l)$ respectively.
Also, note that the action $driveAndTryFix$ and fluents $At_{HL}(l)$, $Spare_{HL}(l)$, $Road_{HL}(o,d)$, and $Flat_{HL}$ are mapped as before.
We can show that $\D_h^{tt+}$ is a sound abstraction of $\D_l^{tt+}$ relative to refinement mapping $m^{tt+}$.

As indicated above, we have complete information at the low level and 
a single model $M_l$ of $\D_l^{tt+}$. Now suppose that the
sequence of (executable) low-level system actions
$\vec{a} = drive(11,21,NoFlatTire);drive(21,12, FlatTire)$ has
occurred. Note that in location $12$, no spare tire is available and if the agent wants to move on with her journey, she needs to either call for service or order a new tire.

The inverse mapping allows us to conclude that sequence of 
high-level system actions $\vec{\alpha} = driveAndTryFix(11,21,DrvNoFlat);$ $ driveAndTryFix(21,12,DrvFlat)$ has occurred,
since $m^{-1}_{M_l}(do(\vec{a},S_0)) = \vec{\alpha}$.
Since $\D_h^{tt+} \models At_{HL}(12, do(\vec{\alpha},S_0))$, we can also
conclude at the high level that the agent is now at location $12$.  As well, since
$\D_h^{tt+} \cup \{ Poss(serviceAndFix(12,Succ_{HServ}), do(\vec{\alpha},S_0)) \}$ is
satisfiable, we can conclude that high-level system action
$serviceAndFix(12,Succ_{HServ})$ might occur next.  Analogously, we can also
conclude that high-level system action $buyAndFix(12,Succ_{HBuy})$ might occur
next, in fact we know at the high level that it is executable:
$\D_h^{tt+} \models Poss(buyAndFix(12,Succ_{HBuy}), do(\vec{\alpha},S_0))$.

\section{Discussion}


There has been much previous work on fully/partially observable
nondeterministic planning, e.g., \cite{DBLP:books/daglib/0014222,DBLP:journals/ai/CimattiPRT03,DBLP:conf/aaai/MuiseBM14}, strategy logics such as ATL \cite{DBLP:journals/jacm/AlurHK02}, and strategic reasoning \cite{XiongLiuECAI16,XiongLiuIJCAI16}. But such work assumes that the domain is represented at a single level of abstraction.

In planning, approaches such as hierarchical task networks (HTNs) \cite{DBLP:conf/aaai/ErolHN94} support abstract tasks that facilitate search.
\cite{DBLP:journals/ai/KuterNPT09} propose an algorithm which combines HTN-based search-control strategies with Binary Decision Diagram-based state representations for planning in nondeterministic domains. \cite{DBLP:conf/aips/ChenB21} extend HTN planning with nondeterministic primitive tasks. 
\cite{DBLP:conf/ijcai/BonetGGR17} investigate nondeterministic abstractions for generalized planning, where one looks for a (typically iterative) plan that solves a whole class of related planning problems.
%
\cite{DBLP:conf/ijcai/Cui0L21} applies the abstraction framework of (BDL17) to generalized planning.  They use \Golog programs to represent nondeterministic actions, and our results should make it possible to simplify their framework. 
%

\cite{DBLP:conf/ijcai/BanihashemiGL18} generalized the approach of (BDL17) to deal with agents that may acquire new knowledge during execution, and formalized (piecewise) hierarchical refinements of an agent's ability (i.e., strategy) to achieve a goal. However, this work does not consider nondeterministic actions and is based on a more complex framework involving online executions.

%

Abstraction is important for efficient reasoning and explainability. This paper presents foundational results, i.e.,
a generic framework for abstraction in nondeterministic domains, which can be used for
many reasoning tasks, such as 
planning, execution monitoring, and explanation.
%
One potential practical application area is smart manufacturing. For instance, \cite{DBLP:journals/ai/GiacomoFLPS22} present a ``manufacturing
as a service'' framework where facilities (made up of \emph{concrete} resources) bid to produce products, given an \emph{abstract} product recipe, which is based on an \emph{abstract} information model in the cloud. Our work shows how one can develop an abstract domain model where planning for arbitrary high-level goals is easier and where we have guarantees that a high-level plan/strategy can be refined into a low-level one.
It also allows one to monitor the low-level system by generating high-level descriptions of low-level system executions from which one can reason at the high level about what might happen next. Note that often manufacturing processes depend on the data and objects (parts) they produce and consume; to formalize this aspect, the situation calculus which provides a first-order representation of the state of the processes can indeed be used. Another potential practical application area is business process management, to support execution monitoring, explanation, and failure handling, see e.g., \cite{DBLP:journals/tist/MarrellaMS17}.


As discussed in Section \ref{sec:NDabstraction}, (BDL17) identifies a set of properties that are necessary and sufficient to have a sound/complete abstraction wrt a mapping, which can be used to verify that one has such an abstraction.
This means that in the finite domain/propositional case, verifying that one has a sound abstraction is decidable and one can use theorem proving techniques to do it.
Moreover, when one has complete information and a single model, one can use model-checking techniques.
Bounded basic action theories \cite{DBLP:journals/ai/GiacomoLP16}  constitute an important class of infinite domain theories where verification is decidable.
In the general infinite-domain case, there are sound but incomplete reasoning methods that can be used \cite{DBLP:conf/kr/GiacomoLP10}.

Automatically synthesizing sound/complete abstractions is another interesting problem.
\cite{DBLP:conf/ecai/Luo0LL20} show that one can use the well-explored notion of forgetting (of low-level fluent and action symbols) to automatically obtain a sound and complete high-level abstraction of a low-level BAT \emph{for a given mapping} under certain conditions.  They also show that such an abstraction is computable in the propositional case. However in general, there may be many different abstractions of a low-level theory, each of which may be useful for a different purpose. So defining an abstract language and mapping for a domain is not a trivial problem.  Some human intervention is likely to be required, e.g., the modeler might specify the goals of the abstraction, or which details can be considered unimportant.

In this paper, we have not considered fairness constraints and strong cyclic plans \cite{DBLP:journals/jair/DIppolitoRS18,DBLP:conf/aips/AminofGR20}; this is a topic for future work.
Another interesting direction for future research is to consider a notion of multi-tier planning \cite{DBLP:conf/aips/CiolekDPS20}, where planning may be done using a simple model without unlikely contingencies, but where one can fall back to a more complex model when such contingencies do occur or more robustness is required.
%
Related to this is the
notion of ``best effort strategies'' (in presence of multiple [contradictory] assumptions about the environment) as proposed by \cite{DBLP:conf/kr/AminofGLMR21}. These are agent plans which, for each of the environment specifications individually, realize the agent's goal against a maximal set of environments satisfying that specification.
We are also interested in extending the current framework to deal with partial observability and sensing.


 

\section*{Acknowledgments}
Work supported by the ERC Advanced Grant WhiteMech (No. 834228), by 
the EU ICT-48 2020 project TAILOR (No. 952215), as well as by the 
National Science and Engineering Research Council of Canada.

\bibliographystyle{named}
\bibliography{citations}

\appendix


\section{Proofs}

In this paper, we use the following standard first-order logic
notations: $M,v \models \phi$ means that model $M$ and assignment $v$
satisfy  formula $\phi$ (where  $\phi$ may contain free variables that
are interpreted by $v$); also $v[x/e]$ represents the variable
assignment that is just like $v$ but assigns variable $x$ to entity $e$.


\vspace{1ex}

\noindent
\textbf{Theorem \ref{thm:bisimProperMgeneralizes2agtAct} }
  Suppose that  $M_h \sim_m M_l$, where  $M_h \models \D_h$, $M_l
  \models \D_l \cup \C$, $\D_h$ and $\D_l$ are both NDBATs, and $m$ is
  proper wrt $D_l$.
  Then for any high-level system action sequence $\vec{\alpha}$ and any high-level
      primitive action $A$, we have that:
  \begin{enumerate}
  \item
      if there exist  $s_h$ and $s_h'$ such that
      $M_h,v[s/s_h,s'/s_h'] \models Do(\vec{\alpha},S_0,s) \land
      Do_{ag}(A(\vec{x}),s,s')$,
      then there exist  $s_l$ and $s_l'$ such that $M_l,v[s/s_l,s'/s_l'] \models
      Do(m_s(\vec{\alpha}),S_0,s) \land 
      Do_{ag}(m_a(A(\vec{x})),s,s')$, 
$s_h \sim_m^{M_h,M_l} s_l$, and $s_h' \sim_m^{M_h,M_l} s_l'$.

    \item
      if there exist $s_l$ and $s_l'$ such that $M_l,v[s/s_l,s'/s_l'] \models
      Do(m_s(\vec{\alpha}),S_0,s) \land  Do_{ag}(m_a(A(\vec{x})),s,s')$,
      then there exist $s_h$ and $s_h'$ such that
      $M_h,v[s/s_h,s'/s_h'] \models  Do(\vec{\alpha},S_0,s) \land
      Do_{ag}(A(\vec{x}),s,s')$,
      $s_h \sim_m^{M_h,M_l} s_l$, and $s_h' \sim_m^{M_h,M_l} s_l'$.
    \end{enumerate}

\noindent
\paragraph{Proof:}
(1.)
Assume the antecedent.
We have that there exists  $s_h$ and $s_h'$ such that
      $M_h,v[s/s_h,s'/s_h'] \models Do(\vec{\alpha},S_0,s) \land
      Do_{ag}(A(\vec{x}),s,s')$.
Since $M_h \models D_h$ and $D_h$ is an NDBAT, it satisfies the
reaction existence requirement, so we have that there exists a
reaction $e_h$ such $M_h,v[s/s_h,s'/s_h', e/e_h] \models
Do(A(\vec{x},e),s,s')$.
By Lemma \ref{lem:bisimL2HOffND}, we have that
there exists $s_l$ such that
$M_l,v[s/s_l] \models Do(m_s(\vec{\alpha}),S_0,s)$ and 
$s_h \sim_m^{M_h,M_l} s_l$.
Thus by the definition of $m$-bisimulation, we have that
there exists $s_l'$ such that
$M_l,v[s/s_l,s'/s_l',e/e_h] \models
      Do(m_s(A(\vec{x},e)),s,s')$ and $s_h' \sim_m^{M_h,M_l} s_l'$.
Since $m$ is proper wrt $D_l$, then it follows that
$M_l,v[s/s_l,s'/s_l'] \models  Do_{ag}(m_a(A(\vec{x})),s,s')$.
\\
(2.)
Assume the antecedent.
We have that there exists
$s_l$ and $s_l'$ such that $M_l,v[s/s_l,s'/s_l'] \models
Do(m_s(\vec{\alpha}),S_0,s) \land  Do_{ag}(m_a(A(\vec{x})),s,s')$.
Since $m$ is proper wrt $D_l$, it follows that there exists $e_h$ such
that $M_l,v[s/s_l,s'/s_l',e/e_h] \models  Do(m_s(A(\vec{x},e)),s,s')$.
By Lemma \ref{lem:bisimL2HOffND}, we have that
there exists $s_h$ such that $M_h,v[s/s_h] \models
Do(\vec{\alpha},S_0,s)$ and
 $s_h \sim_m^{M_h,M_l} s_l$.
Thus by the definition of $m$-bisimulation, we have that there exist
$s_h'$ such that
      $M_h,v[s/s_h,s'/s_h',e/e_h] \models  Do(A(\vec{x},e),s,s')$
and $s_h' \sim_m^{M_h,M_l} s_l'$.
Since $M_h \models D_h$ and $D_h$ is an NDBAT, it satisfies the
reaction independence requirement, so we have that
$M_h,v[s/s_h,s'/s_h'] \models Do_{ag}(A(\vec{x}),s,s')$.
\qed

\noindent
\textbf{Proposition \ref{prop:ProperRefEx}}  
NDBAT refinement mapping $m^{tt}$ is \emph{proper} wrt $\D_l^{tt}$.

 

\noindent
\paragraph{Proof:}
For the HL action $driveAndTryFix(o,d)$, we need to show that for
every high-level system action sequence $\vec{\alpha}$, we have that:

\begin{small}  
\[\begin{array}{l}
\D_l^{tt} \cup \C \models \forall s. (Do(m^{tt}_s(\vec{\alpha}),S_0,s) \limp\\
\ \forall o,d,s'. (Do_{ag}(m^{tt}_a(driveAndTryFix(o,d)),s,s') \equiv \\
\exists r_h. Do(m^{tt}_s(driveAndTryFix(o,d,r_h)),s,s')))
\end{array}\]
\end{small}%

\noindent
($\Rightarrow$) 

Take arbitrary $M_l^{tt}$ and $v$ such that $M_l^{tt} \models \D_l^{tt} \cup \C $.
Assume that $M_l^{tt},v \models Do(m^{tt}_s(\vec{\alpha}),S_0,s) \land Do_{ag}(m^{tt}_a(driveAndTryFix(o,d)),s,s')$. 
Hence, we have that there exists $s_i$ such that  $M_l^{tt},v[s''/s_i] \models Do(m^{tt}_s(\vec{\alpha}),S_0,s) \land Do_{ag}(drive(o,d),s,s'')$.

By definition of $Do_{ag}$, we have that $M_l^{tt},v[s''/s_i] \models Do(m^{tt}_s(\vec{\alpha}),S_0,s) \land \exists r_l.Poss(drive(o, d, r_l),s) \land s''=do(drive(o,d,r_l),s)$.

\noindent
We can consider the following cases:

\begin{itemize}
	\item  case 1: Suppose $M_l^{tt},v[s''/s_i] \models \neg Flat_{LL}(s'')$. Then we have that
				 $M_l^{tt},v[s''/s_i] \models \exists
                                 \delta',r_l,r_h.$\\ $Trans(m^{tt}_a(driveAndTryFix(o,d)),s,\delta',s'')
                                 \land 
Poss(drive(o, d, r_l),s) \land s''=do(drive(o,d,r_l),s)  \land r_l =
NoFlatTire \land r_h=DrvNoFlat \land \land Final(\delta',s'') \land s'' = s'$. Hence we have that  $M_l^{tt},v \models \exists r_h. Do(m^{tt}_s(driveAndTryFix(o,d,r_h)),s,s')$.

	\item  case 2: Suppose $M_l^{tt},v[s''/s_i] \models Flat_{LL}(s'')$. Then $r_l$ must be $FlatTire$.

	 \begin{itemize}
		 \item case 2.1: Suppose $M_l^{tt},v[s''/s_i] \models \neg Spare(d,s'')$.  Then
                     we have that  $M_l^{tt},v[s''/s_i] \models
                     \exists \delta',r_l,r_h.$ $
                     Trans(m^{tt}_a(driveAndTryFix(o,d)),s,\delta',s'') \land
Poss(drive(o, d, r_l),s) \land s''=do(drive(o,d,r_l),s)  \land r_l =
FlatTire \land r_h= DrvFlat \land Final(\delta',s'') \land s'' = s'$.  
										 Hence we have that  $M_l^{tt},v \models \exists r_h. Do(m^{tt}_s(driveAndTryFix(o,d,r_h)),s,s')$.

		 \item case 2.2: Suppose $M_l^{tt},v[s''/s_i] \models Spare(d,s'')$. In this case,
                     we have that there exists $s_j$ such that $M_l^{tt},v[s''/s_i,s'''/s_j] \models Do_{ag}(fixFlatTire(d),s'',s''')$. By  
										 definition of $Do_{ag}$, we have that $M_l^{tt},v[s''/s_i,s'''/s_j] \models \exists
                      r_l'.Poss(fixFlatTire(d, r_l'),s'') \land r_l'=Success_{LF}$. 
											Thus
                                                                                        we
                                                                                        have
                                                                                        that
                                                                                        $M_l^{tt},v[s''/s_i,s'''/s_j]
                                                                                        \models
                                                                                        \exists
                                                                                        \delta',
                                                                                        \delta'',
                                                                                        r_l,r_l',r_h.$ $
                                                                                        Trans(m^{tt}_a(driveAndTryFix(o,d)),s,\delta',s'')
                                                                                        \land
                                                                                        s''
                                                                                        =
                                                                                        do(drive(o,d,r_l),s)
                                                                                        \land
                                                                                        r_l=FlatTire \land
Trans(\delta',s'',\delta'',s''') \land
											s'''=do(fixFlatTire(d,
                                                                                        r_l'),s'')
                                                                                        \land
                                                                                        r_l'=
                                                                                        Success_{LF}
                                                                                        \land
                                                                                        r_h=DrvFlatFix
                                                                                        \land
                                                                                        Final(\delta'',s''')
                                                                                        \land
                                                                                        s'''=s'$.
											Hence we have that $M_l^{tt},v \models \exists r_h. Do(m^{tt}_s(driveAndTryFix(o,d,r_h)),s,s')$.
											

	 \end{itemize}
	
\end{itemize}

\noindent
($\Leftarrow$)\\
Following similar reasoning as above, there are 3 cases where
$Do(m_s^{tt}(driveAndTryFix(o,d,r_h)),s,s')$ holds for some $r_h$.
It is easy to show that $Do_{ag}(m_a^{tt}(A(\vec{x})),s,s')$ holds for
all the $s'$ that are produced.
\\[1ex]
For the other high level actions, the result follows trivially.
\qed

\noindent
\textbf{Proposition \ref{prop:SoundCompEx}} $\D^{tt}_h$ is a sound and complete abstraction of $\D^{tt}_l$ wrt $m^{tt}$.

\noindent
\paragraph{Proof:}
To prove $\D^{tt}_h$ is a sound abstraction of $\D^{tt}_l$, we use Theorem \ref{thm:verifySound}: 
\\
(a) $\D^{tt}_{l(S_0)} \cup \D^{tt}_{l(ca)} \cup \D^{tt}_{l(coa)} \models m^{tt}_f(\phi)$, for all
$\phi \in \D^{tt}_{h(S_0)}$ as $\D^{tt}_{l(S_0)}$ entails all refinements of the fluents
 $Road_{HL}$, $Spare_{HL}$, $Visited_{HL}$, $At_{HL}$, $Dest_{HL}$, $Flat_{HL}$ that  $\D^{tt}_{h(S_0)}$ contains.
\\
\\
(b)
For the high-level system action $\mathit{driveAndTryFix}(o,d,r_h)$ ($\mathit{driveAndTryFix}$ abbreviated as $\mathit{dtf}$), we need to show that for any sequence of high-level system actions $\vec{\alpha}$: 

\begin{small}
\[\begin{array}{l}
\D^{tt}_l \cup \C \models Do(m^{tt}_s(\vec{\alpha}),S_0,s) \limp\\
\quad \forall o,d,r_h. (m^{tt}_f(o \neq d \land At_{HL}(o,s) \land Road_{HL}(o,d,s) \land \\
\quad \neg Flat_{HL}(s) \land (r_h=DrvNoFlat \lor {}\\
\quad \neg Spare_{HL}(d,s)) \land r_h=DrvFlat \lor {}\\
\quad Spare_{HL}(d,s) \land r_h=DrvFlatFix))\\
\quad \equiv \exists s'. Do(m^{tt}_s(\mathit{dtf}(o,d,r_h)),s,s'))
\end{array}\]
\end{small}%

\noindent i.e.,

\begin{small}
\[\begin{array}{l}
\D^{tt}_l \cup \C \models Do(m^{tt}_s(\vec{\alpha}),S_0,s) \limp\\
\forall o,d,r_h. (o \neq d \land At_{LL}(o,s) \land Road_{LL}(o,d,s) \land \\
\quad \neg Flat_{LL}(s) \land (r_h=DrvNoFlat \lor {}\\
\quad \neg Spare_{LL}(d,s) \land r_h=DrvFlat \lor {}\\
\quad Spare_{LL}(d,s) \land r_h=DrvFlatFix))\\
\quad \equiv \exists s'. Do((\pi r_l.drive(o, d, r_l); \\
\quad \mbox{\textbf{if} } \neg Flat_{LL} \mbox{ \textbf{then} } r_h= DrvNoFlat?; 
\mbox{\textbf{ else}} \\ 
\quad \mbox{\textbf{if}} \neg Spare_{LL}(d) \mbox{ \textbf{then} } r_h= DrvFlat?;
\mbox{\textbf{else} } \\
\quad fixFlatTire(d, Success_{LF}); r_h= DrvFlatFix? \mbox{\textbf{endIf}}\\
\quad \mbox{\textbf{endIf}}),s,s')).
\end{array}\]
\end{small}%

($\Rightarrow$) Assume the antecedent. We have that $o \neq d \land At_{LL}(o,s)  \land  Road_{LL}(o,d,s) \land 
\neg Flat_{LL}(s)$ is the precondition of $\pi r. drive(o,d,r)$. In case $r_h=DrvNoFlat$, we have that after execution of this action $\neg Flat_{LL}(do(drive(o,d,r_l),s))$ and $r_l=NoFlatTire$. If $r_h=DrvFlat$, and $Flat_{LL}(do(drive(o,d,r),s)) \land \neg Spare_{LL}(d,do(drive(o,d,r_l),s))$ holds, then $r_l=FlatTire$. Otherwise when $r_h=DrvFlatFix$, we have $Spare_{LL}(d,do(drive(o,d,r_l),s)) \land Flat_{LL}(do(drive(o,d,r_l),s)) $, and together with $At_{LL}(d,do(drive(o,d,r_l),s))$ these constitute the preconditions of $fixFlatTire(d,Success_{LF})$, and $r_l=FlatTire$. 

\noindent
($\Leftarrow$) Following similar reasoning as above.
\\
\\
For the high-level action $Wait_{HL}$ the result can be shown to hold similar to above.
\\ 
\\
(c) 
For the high-level action $driveAndTryFix$ (abbreviated as $\mathit{dtf}$), we must show that:

\begin{small}
\[\begin{array}{l}
\D^{tt}_l \cup \C \models Do(m^{tt}_s(\vec{\alpha}),S_0,s) \limp\\
\forall  o,d,r_h,s'. (Do(m^{tt}_s(\mathit{dtf}(o,d,r_h)),s,s') \limp \\
\quad \bigwedge_{F_i \in \F^h}  \forall \vec{y} (m^{tt}_f(\phi^{ssa}_{F_i,dtf}(\vec{y},o,d,r_h))[s] \equiv m^{tt}_f(F_i(\vec{y}))[s'])).
\end{array}\]
\end{small}%
\\
\\
\noindent
\textbf{$At_{HL}.$}  For the high-level fluent $At_{HL}$ we must show that:

\begin{small}
\[\begin{array}{l}
\D^{tt}_l \cup \C \models Do(m^{tt}_s(\vec{\alpha}),S_0,s) \limp\\
\forall  o,d,r_h,s'. (Do(m^{tt}_s(\mathit{dtf}(o,d,r_h)),s,s') \limp \\
\quad \forall l(m^{tt}_f(\exists o',r'_h. dtf(o,d,r_h) = dtf(o',l,r'_h) \lor \\
\quad At_{HL}(l)[s] \land \forall d',r''_h. dtf(o,d,r_h) \neq \mathit{dtf}(l,d',r''_h) )\\
\quad \equiv m^{tt}_f(At_{HL}(l))[s']))
\end{array}\]
\end{small}%

\noindent i.e.,

\begin{small}
\[\begin{array}{l}
\D^{tt}_l \cup \C \models Do(m^{tt}_s(\vec{\alpha}),S_0,s) \limp\\
\forall  o,d,r_h,s'. (Do(m^{tt}_s(\mathit{dtf}(o,d,r_h)),s,s') \limp \\
\quad \forall l((o=o' \land d=l \land r_h=r'_h) \lor \\
\quad At_{LL}(l)[s] \land \neg (o=l \land d=d' \land r_h=r''_h) \\
\quad \equiv At_{LL}(l)[s']))
\end{array}\]
\end{small}%

This follows given the successor state axioms of $At_{LL}$ and refinement of $m^{tt}_s(\mathit{dtf}(o,d,r_h))$. 
\\
\\
\noindent
\textbf{$Flat_{HL}.$}
For the high-level fluent $Flat_{HL}$ we must show that:

\begin{small}
\[\begin{array}{l}
\D^{tt}_l \cup \C \models Do(m^{tt}_s(\vec{\alpha}),S_0,s) \limp\\
\forall  o,d,r_h,s'. (Do(m^{tt}_s(dtf(o,d,r_h)),s,s') \limp \\
\quad m^{tt}_f(\exists o',d'. \mathit{dtf}(o,d,r_h) = \mathit{dtf}(o',d',DrvFlat) \lor Flat_{HL}[s])\\
\quad  \equiv m^{tt}_f(Flat_{HL})[s'])
\end{array}\]
\end{small}%

\noindent i.e.,

\begin{small}
\[\begin{array}{l}
\D^{tt}_l \cup \C \models Do(m^{tt}_s(\vec{\alpha}),S_0,s) \limp\\
\forall  o,d,r_h,s'. (Do(m^{tt}_s(dtf(o,d,r_h)),s,s') \limp \\
\quad (o=o' \lor d=d' \lor r_h=DrvFlat) \lor Flat_{LL}[s]\\
\quad  \equiv Flat_{LL}[s'])
\end{array}\]
\end{small}%

This follows given the successor state axioms of $Flat_{LL}$ and refinement of $m^{tt}_s(\mathit{dtf}(o,d,r_h))$.
\\
\\
\noindent
\textbf{$Visited_{HL}.$} For the high-level fluent $Visited_{HL}$ we must show that:

\begin{small}
\[\begin{array}{l}
\D^{tt}_l \cup \C \models Do(m^{tt}_s(\vec{\alpha}),S_0,s) \limp\\
\forall  o,d,r_h,s'. (Do(m^{tt}_s(driveAndTryFix(o,d,r_h)),s,s') \limp \\
\quad \forall l(m^{tt}_f(\exists o',r'_h. \mathit{dtf}(o,d,r_h) = \mathit{dtf}(o',l,r'_h) \lor Visited_{HL}(l)[s])\\
\quad \equiv m^{tt}_f(Visited_{HL}(l))[s']))
\end{array}\]
\end{small}%

\noindent i.e.,

\begin{small}
\[\begin{array}{l}
\D^{tt}_l \cup \C \models Do(m^{tt}_s(\vec{\alpha}),S_0,s) \limp\\
\forall  o,d,r_h,s'. (Do(m^{tt}_s(driveAndTryFix(o,d,r_h)),s,s') \limp \\
\quad \forall l(o=o' \lor d=l \lor r_h=r'_h) \lor Visited_{LL}(l)[s]\\
\quad \equiv Visited_{LL}(l)[s']))
\end{array}\]
\end{small}%

This follows given the successor state axioms of $Visited_{LL}$ and refinement of $m^{tt}_s(\mathit{dtf}(o,d,r_h))$.
\\
\\
For other high-level fluents, the result follows easily as $m^{tt}_s(driveAndTryFix)$ does not affect their refinements.
\\
\\
Note that the high-level action $Wait_{HL}$ does not affect any high-level fluents.

\paragraph{}
To prove $\D^{tt}_h$ is a complete abstraction of $\D^{tt}_l$, we use Theorem \ref{thm:verifySoundComplete}: 
\\
Since we have complete information at both HL and LL NDBAT theories, we have a single model at each level, namely $M_h$ and $M_l$ and as shown in part ($a$) above, we have that $S_0^{M_h} \sim_m^{M_h,M_l} S_0^{M_l}$.
\qed

 
\noindent
\paragraph{Lemma \ref{lem:sitSupBisim}}  If $s_h \sim_m^{M_h,M_l} s_l$, then for any high-level
  situation-suppressed formula $\phi$, we have that:
\\[0.5ex]  
\begin{small}  
\hspace*{1em} $M_h,v[s/s_h] \models \phi[s]\ \ \mbox{if and only if}\ \ 
M_l,v[s/s_l] \models m_f(\phi)[s].$
\end{small}%

\noindent
\paragraph{Proof} The result follows easily by induction on the
structure of $\phi$ using the definition of $m$-bisimilar situations. \qed

\begin{lemma} \label{lem:bisimL2HOffND}
If  $M_h \sim_m M_l$, then
for any sequence of high-level system actions $\vec{\alpha}$,
we have that

\begin{small}  
\[\begin{array}{l}
\mbox{if } M_l,v[s/s_l] \models Do(m_s(\vec{\alpha}),S_0,s)
    \mbox{ then there exists } s_h \mbox{ such that}\\
    \qquad  M_h,v[s/s_h]  \models s = do(\vec{\alpha},S_0) \land
    Executable(s)\\
    \qquad \mbox{and } s_h \sim_m^{M_h,M_l} s_l\\[1ex]
    \mbox{and } \\[1ex]
\mbox{if }  M_h,v[s/s_h]  \models s = do(\vec{\alpha},S_0) \land
    Executable(s)\\
    \mbox{then } M_l,v \models \exists s. Do(m_s(\vec{\alpha}),S_0,s)
    \mbox{ and for all }  s_l \mbox{ such that }\\
    \quad M_l,v[s/s_l] \models Do(m_s(\vec{\alpha}),S_0,s), 
     s_h \sim_m^{M_h,M_l} s_l
\end{array}\]
\end{small}%
\end{lemma}

\noindent
\paragraph{Proof} The result follows easily by induction on the
length of $\alpha$ using the definition of $m$-bisimulation. \qed

\paragraph{Theorem \ref{thm:bisimL2HOffND}} 
  If  $M_h \sim_m M_l$, then 
  for any sequence of high-level system actions $\vec{\alpha}$
  and any high-level situation-suppressed formula $\phi$, 
  we have that

\begin{small}  
\[\begin{array}{l}
  M_l,v \models \exists s'. Do(m_s(\vec{\alpha}),S_0,s') \land
  m_f(\phi)[s'] \hspace{1em} \mbox{ if and only if }\\
  \hspace{4em} M_h,v \models Executable(do(\vec{\alpha},S_0)) \land
  \phi[do(\vec{\alpha},S_0)].
\end{array}\]
\end{small}%

\noindent
\paragraph{Proof} The result follows immediately from Lemma
\ref{lem:bisimL2HOffND} and Lemma \ref{lem:sitSupBisim}.  \qed

\noindent
\paragraph{Theorem \ref{thm:bisimL2HAgtActOffND}} 
  If  $M_h \sim_m M_l$, then 
  for any sequence of high-level agent actions $\vec{\alpha}$ and
  any high-level situation-suppressed formula $\phi$, 
  we have that

\begin{small}  
\[\begin{array}{l}
  M_l,v \models \exists s'. Do_{ag}(m_a(\vec{\alpha}),S_0,s') \land
  m_f(\phi)[s'] \hspace{1em} \mbox{ if and only if }\\
  \hspace{4em} M_h,v \models \exists s'. Do_{ag}(\vec{\alpha},S_0,s') \land
  \phi[s'].
\end{array}\]
\end{small}

\noindent
\paragraph{Proof:}
Assume that $M_h \sim_m M_l$.
We have that

\begin{small}  
\[\begin{array}{l}
    M_h ,v\models \exists s'. Do_{ag}([A_1(\vec{x_1}),\ldots,
    A_n(\vec{x_n})],S_0,s') \land  \phi[s']\\
\hspace{1em} \mbox{ if and only if }\\
    M_h,v \models \exists s'\exists e_1\ldots\exists e_n. \\
    \hspace{1em} 
    Do[A_1(\vec{x_1},e_1), \ldots,
    A_n(\vec{x_n},e_n)],S_0,s') \land  \phi[s']\\
\hspace{5em} \mbox{ by the definitions of $Do_{ag}$ and $Do$ }\\
\hspace{1em} \mbox{ if and only if }\\
    M_l,v \models \exists s'\exists e_1\ldots\exists e_n.\\
    \hspace{1em} 
    Do(m_s([A_1(\vec{x_1},e_1), \ldots,
    A_n(\vec{x_n},e_n)]),S_0,s') \land
    m_f(\phi)[s'] \\
\hspace{5em} \mbox{ by Theorem \ref{thm:bisimL2HOffND} }\\

\hspace{1em} \mbox{ if and only if }\\
    M_l,v \models \exists s'. Do_{ag}(m_a([A_1(\vec{x_1}), \ldots,
    A_n(\vec{x_n})]),S_0,s') \\
\hspace{5.5em}     {} \land
    m_f(\phi)[s'] \\
    \hspace{5em} \mbox{ since $m$ is a proper mapping wrt $\D_l$ }
  \end{array}\]
\end{small} %
\qed

For our running example, we can show that the implementation of all
high-level actions inevitably terminate:

\noindent 
\paragraph{Proposition \ref{prop:NecTerm}} 
NDBAT $\D_l^{tt} $ and mapping $m^{tt}$ satisfy  Constraint \ref{cstr:HLactionsNecTerminate}.
%
%

\noindent
\paragraph{Proof}
For the HL action $driveAndTryFix(o,d)$,
we need to show that
for every high-level system action sequence $\vec{\alpha}$,  all executions of $m_a^{tt}(driveAndTryFix(o,d))$ terminate:

 \begin{small}
   \[\begin{array}{l}
     \D_l^{tt} \cup \C \models \forall s. (Do(m_s^{tt}(\vec{\alpha}),S_0,s) \limp\\
\hspace{1.5em} \forall o,d.(\exists s'.Do_{ag}(m_a^{tt}(driveAndTryFix(o,d),s,s') \supset \\
 \hspace{2.5em} InevTerminates(m_a^{tt}(driveAndTryFix(o,d)),s)))
   \end{array}\]
 \end{small}%

 \noindent
Take arbitrary $M_l^{tt}$ and $v$ such that $M_l^{tt} \models \D_l^{tt} \cup \C $.
Assume that $M_l^{tt},v \models Do(m_s^{tt}(\vec{\alpha}),S_0,s) \land
Do_{ag}(m_a^{tt}(driveAndTryFix(o,d)),s,s')$. Hence, we have that there
exists $s_i$ such that  $M_l^{tt},v[s''/s_i] \models Do(m_s^{tt}(\vec{\alpha}),S_0,s) \land Do_{ag}(drive(o,d),s,s'')$.

By definition of $Do_{ag}$, we have that $M_l^{tt},v[s''/s_i] \models Do(m_s^{tt}(\vec{\alpha}),S_0,s) \land \exists r_l.Poss(drive(o, d, r_l),s) \land s''=do(drive(o,d,r_l),s)$.

\noindent
We can consider the following cases:

\begin{itemize}
	\item  case 1: Suppose $M_l^{tt},v[s''/s_i] \models \neg
          Flat_{LL}(s'')$. Then we
          have that $M_l^{tt},v[s''/s_i] \models  \exists \delta'.
          \Trans(m_a^{tt}(driveAndTryFix(o,d)),s,\delta', s') \; \land
          \exists r_l. 
          s'=do(drive(o,d,r_l),s)  \land r_l = NoFlatTire \land
          Final(\delta', s') \land s'' = s'$.
	
	\item  case 2: Suppose $M_l^{tt},v[s''/s_i] \models
          Flat_{LL}(s'')$. Then $r_l$ must be $FlatTire$.
	 
	 \begin{itemize}
		 \item case 2.1: Suppose $M_l^{tt},v[s''/s_i] \models \neg Spare(d,s'')$. This case is similar to case 1 above.
		 \item case 2.2: Suppose $M_l^{tt},v[s''/s_i] \models Spare_{LL}(d,s'')$. In this case,
                   we have that there exists $s_j$ such that $M_l^{tt},v[s''/s_i,s'''/s_j] \models Do_{ag}(fixFlatTire(d),s'',s''')$. By definition of $Do_{ag}$, we have that
                   $M_l^{tt},v[s''/s_i,s'''/s_j] \models \exists
                   r_l'.Poss(fixFlatTire(d, r_l'),s'') \land
                   r_l'=Success_{LF}$. 
									 Hence,
                                                                         $M_l^{tt},v[s''/s_i,s'''/s_j]
                                                                         \models
                                                                         \exists
                                                                         \delta',\delta''.$ $
                                                                         \Trans(m_a^{tt}(driveAndTryFix(o,d)),s,\delta',
                                                                         s'')
                                                                         \land
                                                                         \exists r_l.
                                                                         s''=do(drive(o,d,r_l),s)
                                                                         \land
                                                                         r_l = FlatTire
                                                                         \land
                                                                         \Trans(\delta',s'',\delta'',s''')
                                                                         \land
                                                                         \exists
                                                                         r_l'
                                                                         =
                                                                         Success_{LF}
                                                                         \land 
                                                                         s'''=do(fixFlatTire(d,r_l'),
                                                                         do(drive(o,d,r_l),s))
                                                                         \land
                                                                         Final(\delta'',s''') \land
                                                                         s'=s'''.$
	 \end{itemize}
	
\end{itemize}

\noindent
Hence, all executions of $m_a^{tt}(driveAndTryFix(o,d)$ terminate. We
can show a similar result for the other high-level actions.
\qed

\paragraph{Length of a strategy.}
We define the length $length(f,s)$ of a strategy $f$ in a situation $s$
as the length of its longest branch; formally:

\begin{small}  
  \[\begin{array}{l}
length(f,s) \doteq\\
 \quad   \begin{cases}
      0 & \text{if  } f(s) = stop\\
      max_{Do_{ag}(f(s),s,s')}(length(f,s')) & \text{otherwise}
    \end{cases}  
\end{array}\]
\end{small}%



\noindent
\paragraph{Theorem \ref{thm:strongPlan}} 
  If  $M_h \sim_m M_l$ and
  Constraint \ref{cstr:AgtAlwsCanExecuteHLactions} hold,
  then 
  for any high-level system action sequence $\vec{\alpha}$ and any high-level situation-suppressed formula $\phi$, 
  we have that:

\begin{small}  
  \[\begin{array}{l}
      \mbox{if }  M_h,v \models Executable(do(\vec{\alpha},S_0)) \land AgtCanForce(\phi, do(\vec{\alpha},S_0)) \\
      \mbox{then } M_l,v \models \exists s. Do(m_s(\vec{\alpha}),S_0,s)
      \land {} \\
      \hspace{5.9em} \forall s. Do(m_s(\vec{\alpha}),S_0,s) \limp  AgtCanForce(m_f(\phi),s)
\end{array}\]
\end{small}%

\noindent
\paragraph{Proof} By induction on the length of the high-level  strategy
$f_h$ at $do(\vec{\alpha},S_0)$.\\[1ex]
Assume the antecedent.
It follows that there is a strategy $f_h$ such
that $M_h,v[s/s_h] \models s = do(\vec{\alpha},S_0) \land AgtCanForceBy(\phi,s, f_h)$.
By Lemma \ref{lem:bisimL2HOffND} we have that
$M_l,v \models \exists s. Do(m_s(\vec{\alpha}),S_0,s)$
and for all $s_l$ such that $M_l,v[s/s_l] \models Do(m_s(\vec{\alpha}),S_0,s), 
     s_h \sim_m^{M_h,M_l} s_l$.
     Take an arbitrary such $s_l$.  We show that there exists a 
     low-level strategy $f_l$ such that
     $M_l,v[s/s_l] \models AgtCanForceBy(m_f(\phi),s,f_l)$.
     \\[1ex]
Base case, when $M_h,v[s/s_h] \models length(f_h,s) = 0$:\\
Then we have that $M_h,v[s/s_h] \models f_h(s) = stop \land \phi[s]$
by the definition of $AgtCanForceBy$.
Therefore $M_l,v[s/s_l] \models m_f(\phi)[s]$ must hold by Lemma
\ref{lem:sitSupBisim},
and thus $M_l,v[s/s_l] \models AgtCanForceBy(m_f(\phi),s_l,f_l)$ for any strategy $f_l$ such that
$f_l(s_l) = stop$.
\\[1ex]
Inductive case:\\
Assume the result holds for any high-level  strategy $f_h$ of length at
most $k$.  We show that it must also hold for a high-level strategy $f_h$ of
length $k+1$.
\\
At the high level, by the definition of $AgtCanForceBy$, we have that
there exists an high-level action $A(\vec{t})$ such that 

\begin{small}  
\[\begin{array}{l}
  M_h,v[s/s_h] \models f_h(s) = A(\vec{t})  \land \exists s'. Do_{ag}(A(\vec{t}),s,s') \land {}\\
\quad  \forall s'. Do_{ag}(A(\vec{t}),s,s') \supset AgtCanForceBy(\phi,s',f_h)
  \end{array}\]
\end{small}

\noindent
Since Constraint \ref{cstr:AgtAlwsCanExecuteHLactions} holds,
the agent always knows how to execute high-level atomic
actions to completion at the low level, and thus there exists a low-level strategy $g_l$ such that 
$M_l ,v[s/s_l]\models AgtCanForceBy(m_a(A(\vec{t})),s,g_l)$.
Take an arbitrary $s_l'$ such that $M_l,v[s/s_l,s'/s_l'] \models
Do_{ag}(g_l,s,s')$.
We have that $M_l,v[s/s_l,s'/s_l'] \models Do_{ag}(m_a(A(\vec{t})),s,s')$.
By Constraint \ref{cstr:ProperMapp}, we have that there exists a HL reaction $e_h$ such
that  $M_l,v[s/s_l,s'/s_l',e/e_h] \models Do(m_s(A(\vec{t},e)),s,s')$.
Since $M_h \sim_m M_l$ and $s_h \sim^{M_h,M_l}_m s_l$, there exists
$s_h'$ such that $s_h' \sim^{M_h,M_l}_m s_l'$ and 
$M_h,v[s/s_h,s'/s_h',e/e_h] \models Do(A(\vec{t},e),s,s')$.
The length of $f_h$ at $s_h'$ is at most $k$.
Thus by the induction hypothesis, there exists a low-level strategy
$f_l'$ such that  $M_l,v[s'/s_l'] \models AgtCanForceBy(m_f(\phi),s',f_l')$.
It follows that $M_l ,v[s/s_l] \models AgtCanForceBy(m_f(\phi),s,f_l)$
for $f_l =  f_l' \circ g_l$.
\qed

\noindent
\paragraph{Theorem \ref{thm:agtCanForceProgram}} 
If  $M_h \sim_m M_l$ and Constraint \ref{cstr:AgtAlwsCanExecuteHLactions} holds,
 then for 
any high-level system action sequence $\vec{\alpha}$ and
 any SD \ConGolog\ high-level agent program $\delta$ without
 the concurrent composition construct,
  we have that:

\begin{small}  
  \[\begin{array}{l}
      \mbox{if }  M_h,v \models Executable(do(\vec{\alpha},S_0)) \land
      {}\\
      \hspace{5em} AgtCanForce(\delta, do(\vec{\alpha},S_0)) \\
      \mbox{then } M_l,v \models \exists s. Do(m_s(\vec{\alpha}),S_0,s)
      \land {} \\
      \hspace{5.9em} \forall s. Do(m_s(\vec{\alpha}),S_0,s) \limp  AgtCanForce(m_a(\delta),s)
\end{array}\]
\end{small}%

\noindent
\paragraph{\emph{Proof Sketch}}
The proof of Theorem \ref{thm:agtCanForceProgram} is also
by induction on the length of the high-level  strategy $f_h$ for
executing $\delta$ at $do(\vec{\alpha},S_0)$ and
is very similar to that of the above theorem.
Note that since $\delta$ is SD, does not involve concurrency, and the
high-level atomic actions are all mapped to SD programs, $m_a(\delta)$
must also be SD (we execute a high-level action using the appropriate
strategy  essentially as a procedure call and the high-level program
continues after the call has completed).
\qed

\noindent
In the above result, we have excluded the program $\delta$ from using
concurrency.  Handling it at the high level is not problematic, but
when $\delta$ is mapped to the low level, we may get interleaved
executions of the mapped high-level atomic actions, which can lead to
low-level situations that are not $m$-bisimilar to the high-level ones
(also, the result of mapping $\delta$ may no longer be SD).  Note that
the agent chooses which transitions to follow when it executes the
program (the nondeterminism is angelic), so it can ensure that high-level
actions are not interleaved at the low level.  But we can easily enforce this
by mapping high-level actions to a special construct that executes
them as an atomic block without interleaving \cite{DLLmutexNote04}.
%
Of course, there may be cases where interleaving the execution of the
implementation of high-level actions is useful and does not lead to
problems.  Addressing this left for future work.

\section{Additional Details on \ConGolog Semantics}

For \emph{system} actions, definitions of $Trans$ and $Final$ for the \ConGolog constructs are as in
\cite{DBLP:conf/kr/GiacomoLP10}:\footnote{Note that since $Trans$ and  $Final$ take programs (that include test of formulas) as arguments, this requires encoding formulas and programs as terms; see
 \cite{DBLP:journals/ai/GiacomoLL00} for details.}

\begin{small}
\[\begin{array}{l}
Trans(\alpha,s,\delta',s') \equiv {}\\
\qquad
s'=do(\alpha,s) \land \Poss(\alpha,s) \land \delta'=True? \\[0.5ex]
Trans(\varphi?,s,\delta',s') \equiv False \\[0.5ex]
Trans(\delta_1; \delta_2,s,\delta',s') \equiv{} \\[0.5ex]
\qquad
	Trans(\delta_1,s,\delta_1',s') \land \delta'=\delta_1';\delta_2 \lor{}\\
\qquad
		Final(\delta_1,s) \land Trans(\delta_2,s,\delta',s') \\[0.5ex]
Trans(\delta_1 \ndet \delta_2,s,\delta',s') \equiv{} \\
\qquad
	Trans(\delta_1,s,\delta',s') \lor Trans(\delta_2,s,\delta',s') \\[0.5ex]
Trans(\pi x.\delta, s, \delta',s') \equiv 
	\exists x. Trans(\delta,s,\delta',s')\\[0.5ex]
Trans(\delta^*, s, \delta',s') \equiv
	Trans(\delta,s,\delta'',s')\land \delta'=\delta'';\delta^*\\
Trans(\delta_1 \conc \delta_2,s,\delta',s') \equiv{} \\ 
\qquad
	Trans(\delta_1,s,\delta_1',s')  \land \delta'=\delta_1'\conc\delta_2 \lor {}\\
\qquad
		Trans(\delta_2,s,\delta_2',s')  \land \delta'=\delta_1\conc\delta_2' 
		\\
		\\

Final(\alpha,s) \equiv False \\
Final(\varphi?,s) \equiv \varphi[s] \\
Final(\delta_1; \delta_2,s) \equiv 
        Final(\delta_1,s) \land Final(\delta_2,s)\\
Final(\delta_1 | \delta_2,s) \equiv 
	Final(\delta_1,s) \lor Final(\delta_2,s) \\
Final(\pi x.\delta, s) \equiv
	\exists x. Final(\delta,s)\\
Final(\delta^*, s) \equiv True\\
Final(\delta_1 \conc \delta_2,s) \equiv 
       Final(\delta_1,s) \land Final(\delta_2,s)
\end{array}\]
\end{small}%
These are in fact the usual ones
\cite{DBLP:journals/ai/GiacomoLL00}, except that, following
\cite{DBLP:conf/kr/ClassenL08}, the test construct $\varphi?$ does not yield
any transition, but is final when satisfied.  Thus, it is
a \emph{synchronous} version of the original test construct (it does
not allow interleaving).

Also, note that the construct \textbf{if} $\phi$ \textbf{then} $\delta_1$
\textbf{else} $\delta_2$ \textbf{endIf} is defined as $[\phi?;
\delta_1 ] \mid [\neg \phi?; \delta_2]$. The while-loop construct, 
\textbf{while} $\phi$ \textbf{do} $\delta$ \textbf{endWhile}
is defined as $(\phi: \delta)^*; \neg \phi?$

\paragraph{}
For primitive \emph{agent} actions, we have $Final(A(\vec{x},s)) \equiv False$ as usual, while 

\begin{small}
\[\begin{array}{l}
Trans(A(\vec{x}),s,\delta',s') \equiv {}\\
\qquad
	\exists e. \Poss(A(\vec{x},e),s) \land \delta'= \epsilon \land s'=do(A(\vec{x},e),s) 
\end{array}\]
\end{small}%

\noindent 
where $\epsilon$ is an abbreviation for $True?$. 

This reflects the fact that agent actions are nondeterministic
and their outcome depends on the environment reaction.
This results in a synchronous test construct that does not allow
interleaving (every transition involves the execution of an action).

The definitions of Trans and Final for the other \ConGolog
constructs are as above.

\end{document}